\documentclass[fleqn,10pt]{wlscirep}
\usepackage[utf8]{inputenc}
\usepackage[T1]{fontenc}
\usepackage{array} 
\usepackage{graphicx} 
\usepackage{subfigure} 
\usepackage{subcaption} 
\usepackage{tabularx}
\usepackage{rotating} 

\title{Topological Components in a Community Currency Network}

\author[1, 2, *]{Teodoro Criscione}
\affil[1]{Central European University, Department of Network and Data Science, Vienna, 1100, Austria}
\affil[2]{Freiburg Institute For Basic Income Studies, University of Freiburg, Freiburg im Breisgau, 79085, Germany}

\affil[*]{criscione\_teodoro@phd.ceu.edu}


\begin{abstract}
Transaction data from digital payment systems can be used to study economic processes at such a detail that was not possible previously. Here, we analyse the data from Sarafu token network, a community inclusion currency in Kenya. During the COVID-19 emergency, the Sarafu was disbursed as part of a humanitarian aid project. In this work, the transactions are analysed using network science. A topological categorisation is defined to identify cyclic and acyclic components. Furthermore, temporal aspects of circulation taking place within these components are considered. The significant presence of different types of strongly connected components as compared to randomized null models shows the importance of cycles in this economic network. Especially, indicating their key role in currency recirculation. In some acyclic components, the most significant triad suggests the presence of a group of users collecting currency from accounts active only once, hinting at a misuse of the system. In some other acyclic components, small isolated groups of users were active only once, suggesting the presence of users only interested in trying out the system. The methods used in this paper can answer specific questions related to user activities, currency design, and assessment of monetary interventions. Our methodology provides a general quantitative tool for analysing the behaviour of users in a currency network.
\end{abstract}
\begin{document}

\flushbottom
\maketitle
\thispagestyle{empty}

\section{Introduction}  \label{sec:introduction}

The transactions in a payment system span a directed, weighted, temporal network, where the nodes are the subjects of the system and the timestamped directed weighted links correspond to those transactions. In this work, the transactions are temporally-aggregated into weighted directed links to study the network topology. Nonetheless, some temporal aspects of circulation will be analysed by considering the timestamped transactions.

This work focuses on Sarafu token network, a digital community currency system used as a payment system in Kenya and organised by the non-profit organisation Grassroots Economics~\cite{mattssonSarafu}. Since the end of 2021, Sarafu network changed its structure and internal policies by integrating also a producer voucher credit system~\cite{ruddick_sarafu_2023}. Nonetheless, here we deal with data from the period when Sarafu token network was used as part of an emergency cash transfer program during the COVID-19 crisis~\cite{ruddick_sarafu_2021}. This humanitarian aid campaign called \textit{"Community Inclusion Currency"} was co-designed with the Kenyan Red Cross. A cash transfer program is used in emergency contexts to transfer money or vouchers to people in need which allow them to buy goods and services. A \textit{"Community Inclusion Currency"} is a specific type of local voucher system used for those humanitarian cash transfer programs, which can be used only in a predefined geographic region or within a local network of participants. In fact, it is argued that, once this local voucher is issued and its recirculation is bounded to a defined geographic area, it could also boost local development whenever the increase of demand in goods and services meets the unused productive capacity of the region~\cite{ussher_complementary_2021}. Nonetheless, a very limited amount of quantitative studies analysed the effectiveness of local voucher systems in cash transfer programs. Their effect on local economy is largely unexplored, but the digitisation of voucher and currency systems open new frontiers to this field of research.

Only in the recent years, a limited number of studies tried to assess the economic impact of digital community currency systems using network analysis. However, many aspects on the topology and the dynamics of payment systems are left to be explored. In this paper, we investigate what the topology of the network can tell us about the usage of the currency. In particular, the currency is studied by identifying cyclic and acyclic components, and their currency recirculation. This technique can help us to distinguish users who engages with the system in different ways. For instance, we can identify users who were fully committed to use the system regularly by both buying and selling, but we can also identify users who followed different strategies. In fact, users who do not engage as expected with the rest of the currency network are an important signal to be considered in assessing the system. Identifying these users and understanding their strategies is a vital aspect for the management of a currency system. Understanding user engagement is especially important in the assessment of \textit{"Community Inclusion Currency"} systems, which are designed for cash transfer programs. Finally, the methods developed in this paper could be also extended for the evaluation of similar monetary interventions, whenever the transaction data are available in a digital format. \\

A \textit{"Community Inclusion Currency"} is a specific type of community currency system. A community currency system is a payment system circulating in a limited geographic region in parallel to the official currency. In particular, a community currency is not enforced by the state but rather based on agreements among members of the community~\cite{blanc_2011, blanc_2012, gomez_book_2018, gomez_dini_2016, greco_money_2001, greco_2013}. Community currency systems have been explored as innovative methods for social and humanitarian interventions which could induce endogenous local development, empower local communities, and at the same time deliver humanitarian aid~\cite{zeller_economic_2020, ussher_complementary_2021, gomez_book_2018, martin_belmonte_crisis_2021, nakazato_empirical_2012, ruddick_complementary_2015, fare_freitas}. Furthermore, their counter-cyclic (or macro-stabilizer) effect may have an important role in enhancing the resilience of the local economy~\cite{stodder_2000, stodder_2016, groppa_2013, gelleri_2021, lucarelli_gobbi, martin_belmonte_crisis_2021}. 

Previous studies on digital community currencies used network science techniques for their characterisation. One of the first study was focused on a convertible community currency Tomamae-cho, which was active in Hokkaido for only three months~\cite{kichiji_network_2008}. Their findings confirmed that the transaction network was characterized by a power law decay of the degree distribution, dis-assortative behaviour, and a "small world" feature. The authors found that the ratio of the exponents of in-degree over out-degree distributions decreases with the velocity of currency. They also measure the \textit{network centralisation}\cite{FREEMAN1978} of the transaction graph defined as the \textit{"ratio of the sum of actual difference between the degree centrality of the most central actor and that of all the other actors in the network and the theoretical maximum possible sum of differences in actor degree centrality"} (from p.282, note 15 \cite{kichiji_network_2008}). One of their main findings is that network centralisation is positively correlated with transaction volume. Regarding the Sarafu data used in this paper, the power law decay and dis-assortative behaviour was already confirmed in a previous work~\cite{mattssonCirculation}. In this work, the power law decay is only reported in the data description in Section~\ref{sec:data} and the Supplementary Material (Document A, Section "Data", Figure S2 and Table S1).

In another work, network indicators were suggested to measure the performance of time-banks~\cite{collom2012}. A time-banking system is a specific type of community currency system where time is used as a unit of account and organised as a mutual credit group. In that work~\cite{collom2012}, two sets of performance indicators were applied to the case of Portland West Time Dollar Exchange in Portland, Maine. The first set reported the number of active members, new members joining each month, transaction volume, average transaction volume, and account balance per each user. The second set reported the number of trading partners, the number of reciprocated links, the ego-network density, and the diversity of exchanged services. Particularly, the second set of indicators is aiming at measuring reciprocity and resilience in such a time-based currency. Similarly, in this work, the identified topological components in Sarafu network are analysed separately by looking at some network metrics which can be used to characterise their flow: number of weakly connected components, number of nodes, number of directed links, number of transactions, and volume.

In a further paper, the network analysis primarily focused on detecting central players and identifying a rich-club of prominent users in the RozLEŤSe system active in Brno, Czech Republic~\cite{frankova2014}. The identification of a rich-club was then used to study the resilience of the economic network by implementing a stress test using an experiment in which users have been removed from the system. A more recent work analysed a basic income community currency located in Berlin, called Circles UBI ~\cite{avanzo2023}. In that paper, the Authors split the network into two periods by calculating the characteristic time through the causal fidelity index and studied how the structure of the network changed over time in terms of coreness and prominent users. In this work, the Sarafu data are temporally aggregated into a weighted directed network to study its topology. The identification of key players is not considered in this work. The aim of this work is to identify groups of users who engage with the system in different ways. 

Some recent works were focused on Sardex network, a business-to-business mutual credit system operating in Sardinia, Italy. In the most recent one~\cite{sardex_2024}, the connectivity of the network is analysed over time by looking at the average directed path length, average degree, diameter, clustering coefficient, and average degree centrality. The Authors concluded that the connectivity of the network increased over time. In another work on Sardex network~\cite{iosifidis}, it was found that a statistically significant presence of directed cyclic motifs is beneficial for that payment system. Moreover, they define prominent nodes based on their participation into directed cycles. The findings suggest that the most prominent nodes have in fact a better performance over time. However, their analysis is focused only on static directed simple cycle of length 2, 3, 4, and 5. Inspired by the importance of cycle motifs in a currency network, in this work we explore the functional role of cyclic components (Section \ref{sec:results:top_cat}). Indeed, a cyclic component can include many cycles of different length. Every node in a cyclic component can belong to many different cycles of different length, from cycles of length 2 to cycles with a length equals to the size of the component itself. In this work, the comparison between cyclic and acyclic components is then carried out to show important behavioural differences among their users.

A recent quantitative study on Sarafu network focused on an inverse estimation of transfer velocity and effective balance~\cite{mattssonVelocity}. In their work, the inverse estimation of the transfer velocity is defined as the average holding time of received funds and calculated on a "first-in-first-out" basis~\cite{mattsson_trajectories_2021}. Its findings suggest a high level of geographic and temporal heterogeneity in the usage of the currency. In particular, the transfer velocity and the effective balance generally had a sharp increase in the first half of 2020, but with some variations between urban and rural areas. Another study on Sarafu network analysed a few aspects of currency circulation~\cite{mattssonCirculation}. In particular, in that work Sarafu network appears to be characterised by three main factors: geographic localisation, cycle motifs (of length 2, 3, 4, and 5), and structural correlations. Moreover, the Authors detected key players by using PageRank centrality: savings groups and faith leaders seem to play a key role in the circulation of Sarafu. In another work, the cooperative behaviour of savings groups is analysed through time by using Sankey diagrams~\cite{cooperative_sarafu_2023, temporal_sarafu_2022}. In that work, the network was split into different periods according to the application of restrictions due to COVID-19 crisis. The Authors observed that the role of savings groups increased, especially when the strictest COVID-19 restrictions were put in place. The percentage of transactions from group accounts to users grew from 8\% to 25\%, the sectors of \textit{food} and \textit{shop} gained importance in the same period, and finally, the geographic heterogeneity increased in terms of spending behaviour. 

Unlike those previous works on Sarafu token network, in this paper only the transactions among users are considered (i.e. group accounts are excluded). Instead of considering the velocity of circulation~\cite{mattssonVelocity}, we define and calculate the recirculation time (Section~\ref{sec:methods:recirculation}). Furthermore, like the other aforementioned works~\cite{mattssonCirculation, cooperative_sarafu_2023, temporal_sarafu_2022}, the analysis on circulation is made on a temporally aggregate network of the whole period. Moreover, in this work, three aspects are analysed for each topological component to identify different usage strategies (Section~\ref{sec:methods:topology}): users who made only one operation (Section~\ref{sec:results:one_tx_users}), three-nodes motifs in acyclic components (Section~\ref{sec:results:dag_analysis}), and recirculation time (Section~\ref{sec:results:recirculation}). This is the main difference from previous works which focused on the activity of user and group accounts at a network level~\cite{mattssonCirculation, cooperative_sarafu_2023, temporal_sarafu_2022}. Finally, instead of focusing on cycle motifs (of length 2, 3, 4, and 5) as previous work on Sarafu data~\cite{mattssonCirculation}, in this work cyclic components are considered. As explained before, each cyclic component is defined here as a strongly connected component, where every node can be involved in one or more cycles of different length.

From a theoretical standpoint, the relation between topology and dynamics in economic and financial networks has already been suggested. In network game theory, the degree centrality and the sparseness of the network are proved to be the major causes of inequality, keeping all the other conditions fixed~\cite{cassese_decentralized_2024}. In an agent-based model simulation, nodes with a high level of betweenness centrality on specific trading paths imposed a \textit{mark-up} on their transactions, and therefore, affecting the price formation on the entire network~\cite{cardoso_effect_2020}. In a recent theoretical discussion paper~\cite{Criscione2022Community}, the authors suggested that the presence of directed cycles in a payment system may be theoretically linked to a redistribution of economic power in it, where economic power is defined by degree and betweenness centrality. Another related work presented the possibility of using cycle detection to reduce the need for liquidity in any payment system~\cite{fleischman_liquidity-saving_2020}. This technology is called in many different ways in the literature: \textit{liquidity saving mechanism}~\cite{fleischman_balancing_2020}, \textit{rescontre}~\cite{boerner2017}, debt clearing~\cite{cui2021} or settling\cite{verhoeff2004}, multilateral compensation~\cite{slobodan1992, gazda2001}, netting or net setting~\cite{shafransky2006}. Since it is a multi-object greedy optimisation problem, many different algorithms have been proposed~\cite{fleischman_balancing_2020, slobodan1992, gazda2001, verhoeff2004, shafransky2006, cui2021, schara2018, patcas2011, patcas2014}. Nonetheless, some of them have in common the detection of cycles in a temporally aggregated and weighted graph to set off the outstanding net positions in an obligation network~\cite{gazda2001, shafransky2006, cui2021}. This technique for multilateral compensation using cycle detection was presumably invented by European medieval merchants~\cite{boerner2017, boerner2023}, then it was perfected and adopted by the inter-banking networks in recent times~\cite{bech2001, shafransky2006}. To sum up, the role of cycles in payment systems has been increasingly recognised for its structural functionality. For this reason, this work compares behavioural aspects related to users in cyclic and acyclic components. The findings suggest indeed that the participation (or not) into cycles correspond to different user strategies. In general, a user who participate into a cycle in engaging in both buying and selling, and therefore signalling a full engagement with the economic network.\\

Concluding, the main contributions of this work can be sum up as follows. Firstly, we suggest a topological categorisation for directed networks, which is specifically related to payment systems (Section \ref{sec:results:top_cat}). Indeed, the categorisation of cyclic and acyclic components can help identifying some behavioural dynamics which were left unseen in previous works on the structure of directed networks~\cite{broder_2000, dorogovstev_2001, donato_2008, dorogovstev_2017}. Indeed, the presence of cycles in a payment system was already related to its performance in a previous work~\cite{iosifidis}, while in this work the user behaviour is successfully analysed by using a different but related topological categorisation. Secondly, we study the effect of cycles in a payment system, without any arbitrary limit on their size. Previous studies on cycles in payment systems were focused on the detection of cycles up to length 5~\cite{iosifidis, mattssonCirculation}, but in this work cyclic components are analysed in their integrity. This means that each cyclic component can have cycles from size 2 to a size equals to the length of that component itself. Thirdly, temporal aspects of circulation are analysed using different tools which can help identifying specific behavioural dynamics. For instance, instead of focusing on estimating the transfer velocity of the system~\cite{mattssonVelocity}, in this work the circulation is studied by categorising each recirculation operation and classifying users based on it (Section~\ref{sec:results:recirculation}). Moreover, we found that users who used the system only once can signal important aspects of currency circulation (Section~\ref{sec:results:one_tx_users}). In our knowledge, this approach was never considered before to analyse currency networks. 

Finally, this work is focused on studying the circulation in the economic network, while previous works considered also the financial network (e.g. borrowing, lending) of savings groups~\cite{mattssonCirculation, mattssonVelocity}, or exclusively their financial operations~\cite{cooperative_sarafu_2023}. Beyond the abstract monetary exchange for financial purposes, the analysis of pure economic exchange can reveal information pertaining the flow of real goods and services. Moreover, in the considered period, while the creation of group accounts was somehow supervised by system administrators, the creation of user accounts was not. This means that single users could strategize around the creation of new accounts to get some benefit (Section~\ref{sec:discussion}). This is one of the key aspects behind the research questions tackled in this paper. Indeed, the main research questions unfolded throughout the paper are the following:
\begin{enumerate}
\item \textbf{RQ1}. \textit{What are the most relevant topological components in the network of a payment system?} A topological categorisation is presented to uniquely assign nodes and edges to components based on their systemic functionalities. 
\item \textbf{RQ2}. \textit{How does circulation of currency differ among these topological components?} After having defined those topological components, their temporal behaviour is analysed. In particular, the recirculation of currency and its frequency are analysed, along with one-time usage.  
\item \textbf{RQ3}. \textit{Is there a relationship between human behaviour and the topological components observed in the currency network?} The topological and circulation analyses can help identifying different levels of user engagement. For instance, the significance of certain triads in some topological components can be related to specific behavioural strategies. 
\end{enumerate}

One of the main findings of this paper is that the identified topological categories are shown to be relevant for the study of a payment system (in Section \ref{sec:results:top_cat}). Some of those topological categories can be indeed related to different user strategies. Indeed, the temporal behaviour of users within those components reveal different types and levels of engagement within the economic network (in Section \ref{sec:results:one_tx_users} and \ref{sec:results:recirculation}). Moreover, the significance of some types of triads in acyclic components can confirm the existence of particular strategies (in Section \ref{sec:results:dag_analysis}). The existence of those strategies are also matching the findings of a recent qualitative study~\cite{gaming_sarafu2024} (Section \ref{sec:discussion}). Another important finding is about the relevance of cycle analysis. Indeed, in accordance to recent studies on the role of cycles in payment systems~\cite{iosifidis, mattssonCirculation}, also in this work the cyclic components are found to be relevant for the circulation of currency. Finally, similarly to a previous study on savings groups in Sarafu network~\cite{cooperative_sarafu_2023, temporal_sarafu_2022}, it is possible to conclude that Sarafu token network generally succeeded in stimulating the local economy during the COVID-19 crisis by engaging the majority of its users within its economic network. 

The structure of the paper is described as follows. In the next Section~\ref{sec:methods}, we present the topological characterisation adopted in this paper, also in relation to existing techniques. In Section \ref{sec:data}, the data is described by providing also geographic and economic information of the transaction network. In Section \ref{sec:results}, the topological and temporal analyses are presented (RQ1 and RQ2 are answered). In particular, in Section \ref{sec:results:top_cat} the network is analysed in its predefined topological components by providing information about their statistical significance. In Section \ref{sec:results:one_tx_users}, users who used the system only once are considered in relation to the those topological components. In Section \ref{sec:results:dag_analysis}, the acyclic components are analysed by using a triadic census. In Section \ref{sec:results:recirculation}, users are assigned to different temporal categories according to their speed of recirculation. Finally, in Section \ref{sec:discussion}, the results are interpreted to describe possible user strategies, which are also reported in previous qualitative works (RQ3 is answered).

\section{Background} \label{sec:methods}

\subsection{Network Topology} \label{sec:methods:topology}

The topological categorisation of cyclic components, acyclic components, and single-nodes is used in this work for behavioural investigations on the economic network. Previous studies already analysed the inner structure of directed networks \cite{broder_2000, dorogovstev_2001, donato_2008, dorogovstev_2017}. According to the existing literature, directed networks are generally characterized by a \textit{bow-tie} structure with a largest strongly connected component (\textit{SCC}) as the core where nodes are mutually reachable from each other, a set of nodes which are only sending to the \textit{SCC}, called \textit{IN}-component, and a set of nodes which are only receiving from the \textit{SCC}, called \textit{OUT}-component. Attached to the \textit{IN}- and \textit{OUT}-components there are \textit{tendrils} which can be either sets of nodes that a) can be reached only from the \textit{IN}-component or b) sets of nodes which do not belong to  the \textit{SCC} but can reach nodes of the \textit{OUT}-component. Other groups of nodes, called \textit{tubes}, connect the \textit{IN-}component with the \textit{OUT-}component without passing through the \textit{SCC}. Finally, isolated groups of nodes are just described as \textit{disconnected} components. The \textit{bow-tie} description was successfully applied to study the structure of directed graphs. Especially, it can be used to study some of their properties like their expected size, degree distributions, and resilience to random failures and targeted attacks \cite{dorogovstev_2001, newman_arbitrary_random_2001}. A subsequent work on the web graph reviewed the \textit{bow-tie} structure suggesting a \textit{daisy} shape, where the \textit{IN-} and \textit{OUT-} components are highly fragmented into many \textit{petals}, which are chains of nodes connected with the same component of origin~\cite{donato_2008}. Finally, a more advanced version of the \textit{bow-tie}~\cite{dorogovstev_2017} was recently introduced where \textit{tendrils} and \textit{tubes} are categorised based on their distance from the \textit{IN-} and \textit{OUT-} components. The results of these works have also important implications for the study on the resilience of directed networks. \\

\begin{figure}[!ht]
    \centering
    \subfigure[Cyclic Component]
        {\includegraphics[width=0.3\textwidth]{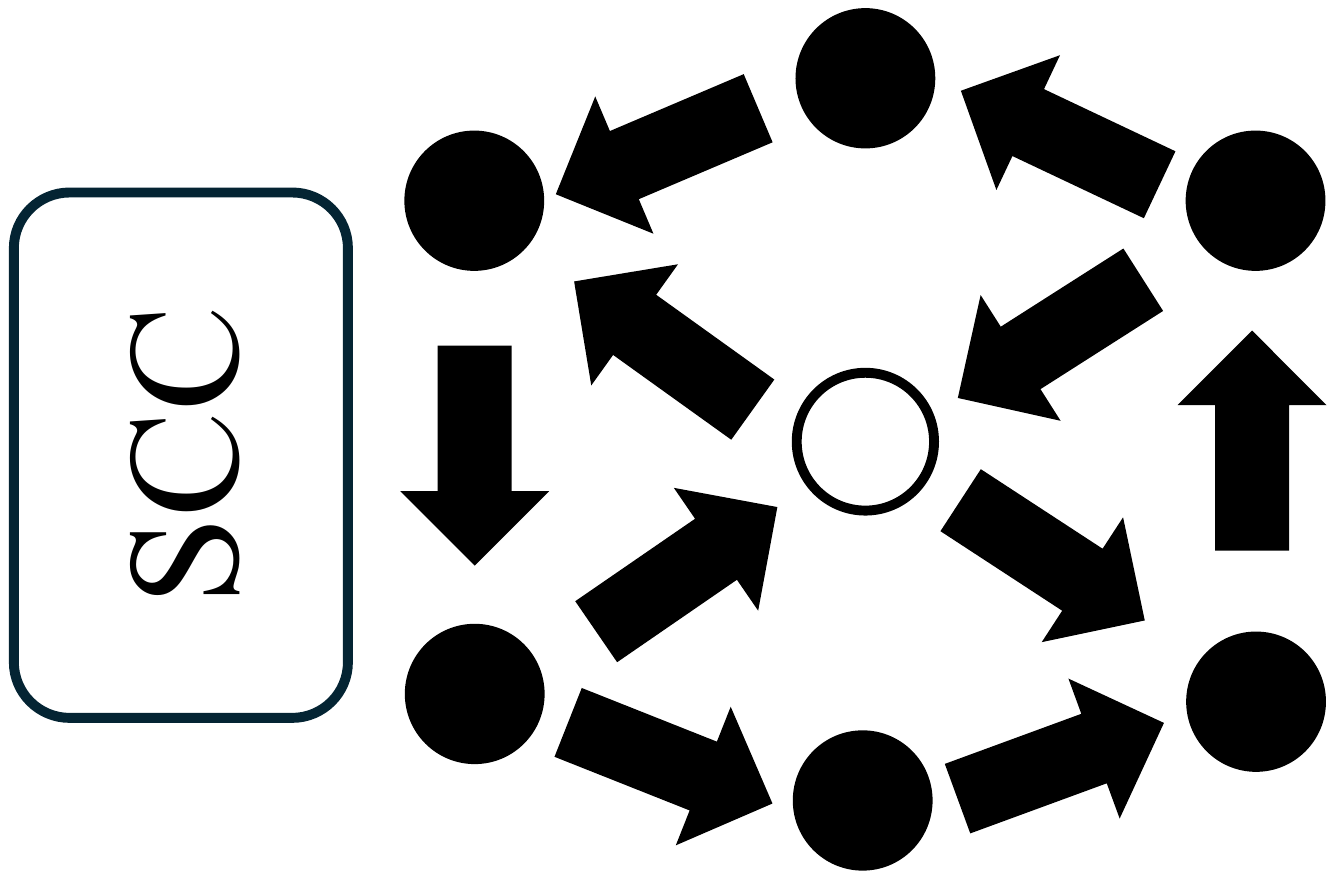}}
    \hspace{0.01\textwidth}
    \subfigure[Acyclic Component]
        {\includegraphics[width=0.3\textwidth]{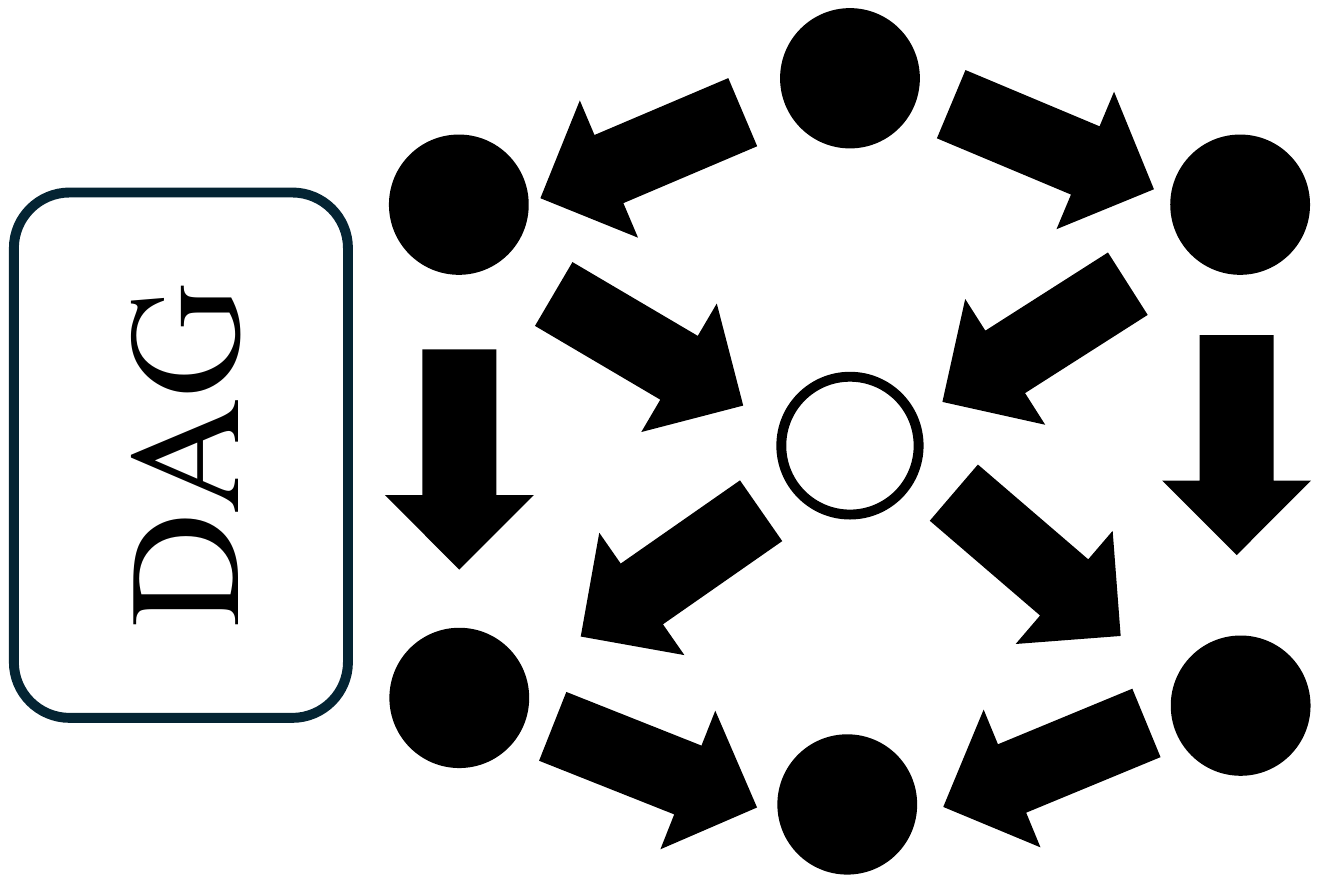}}
    \hspace{0.01\textwidth}
    \subfigure[Single-nodes]
        {\includegraphics[width=0.3\textwidth]{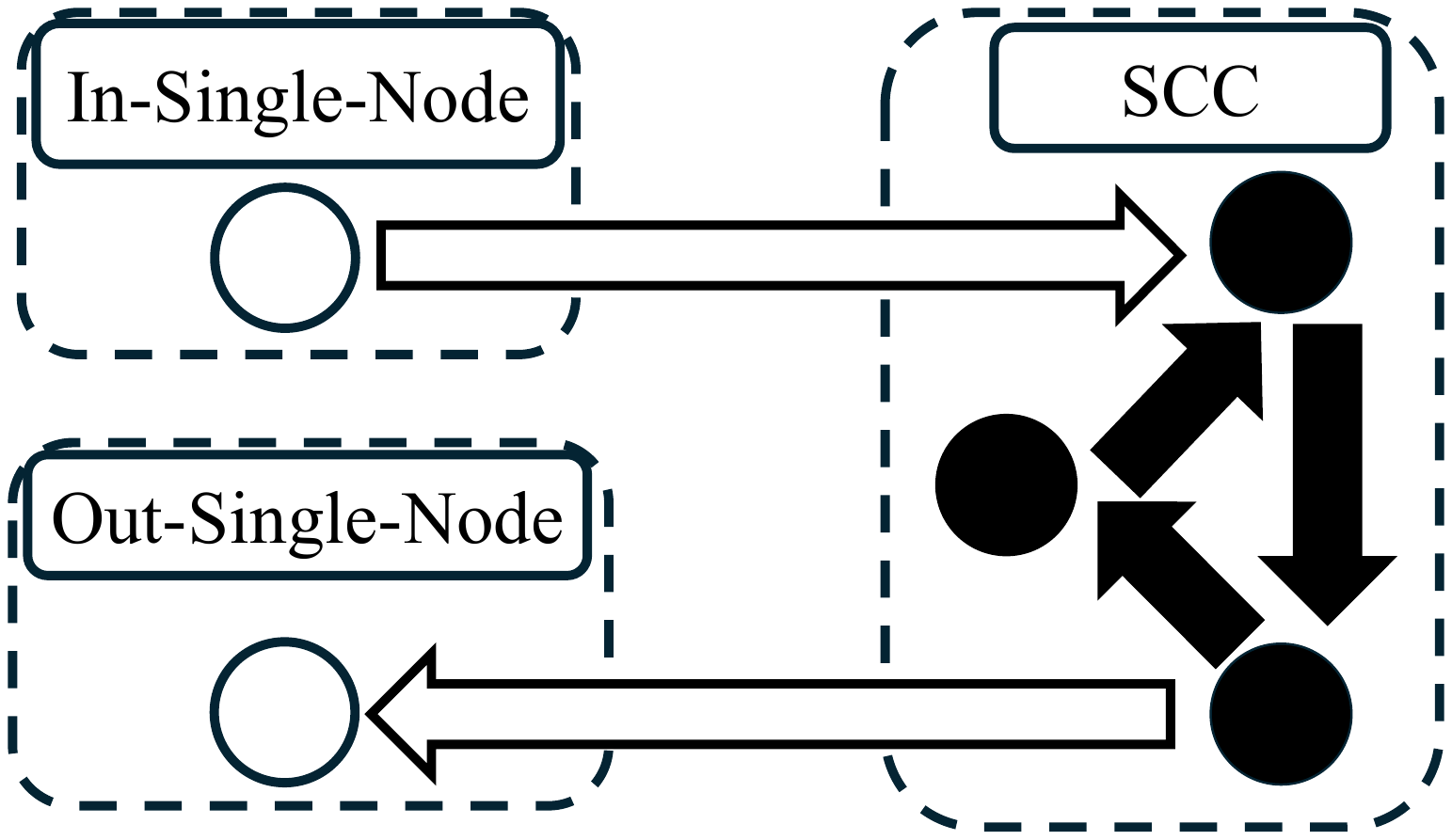}} 
    \\
    \subfigure[Directed link ($\longrightarrow$).]
        {\includegraphics[width=0.3\textwidth]{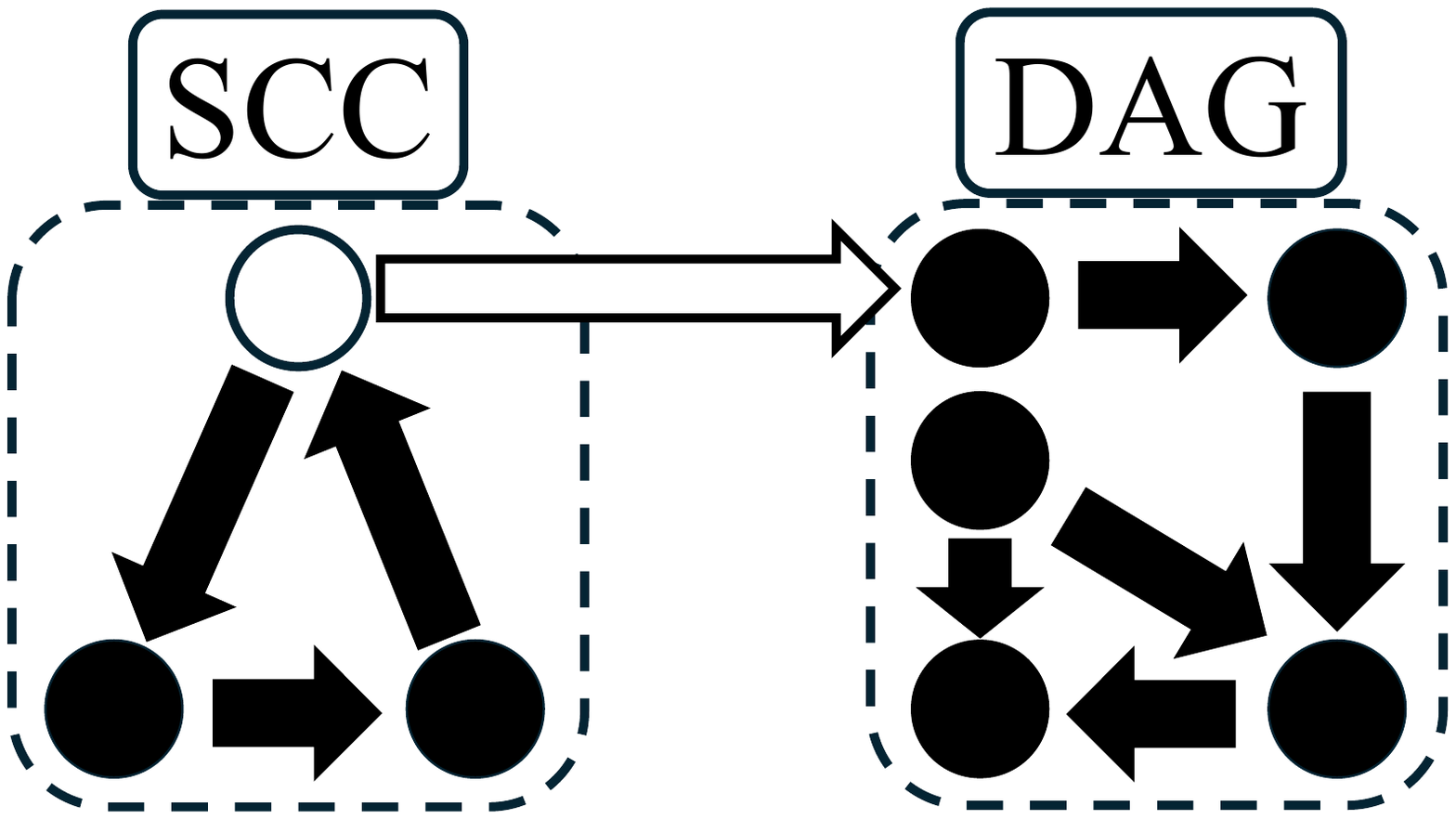}}
    \hspace{0.01\textwidth}
    \subfigure[Double-link ($\leftrightarrows$).]
        {\includegraphics[width=0.3\textwidth]{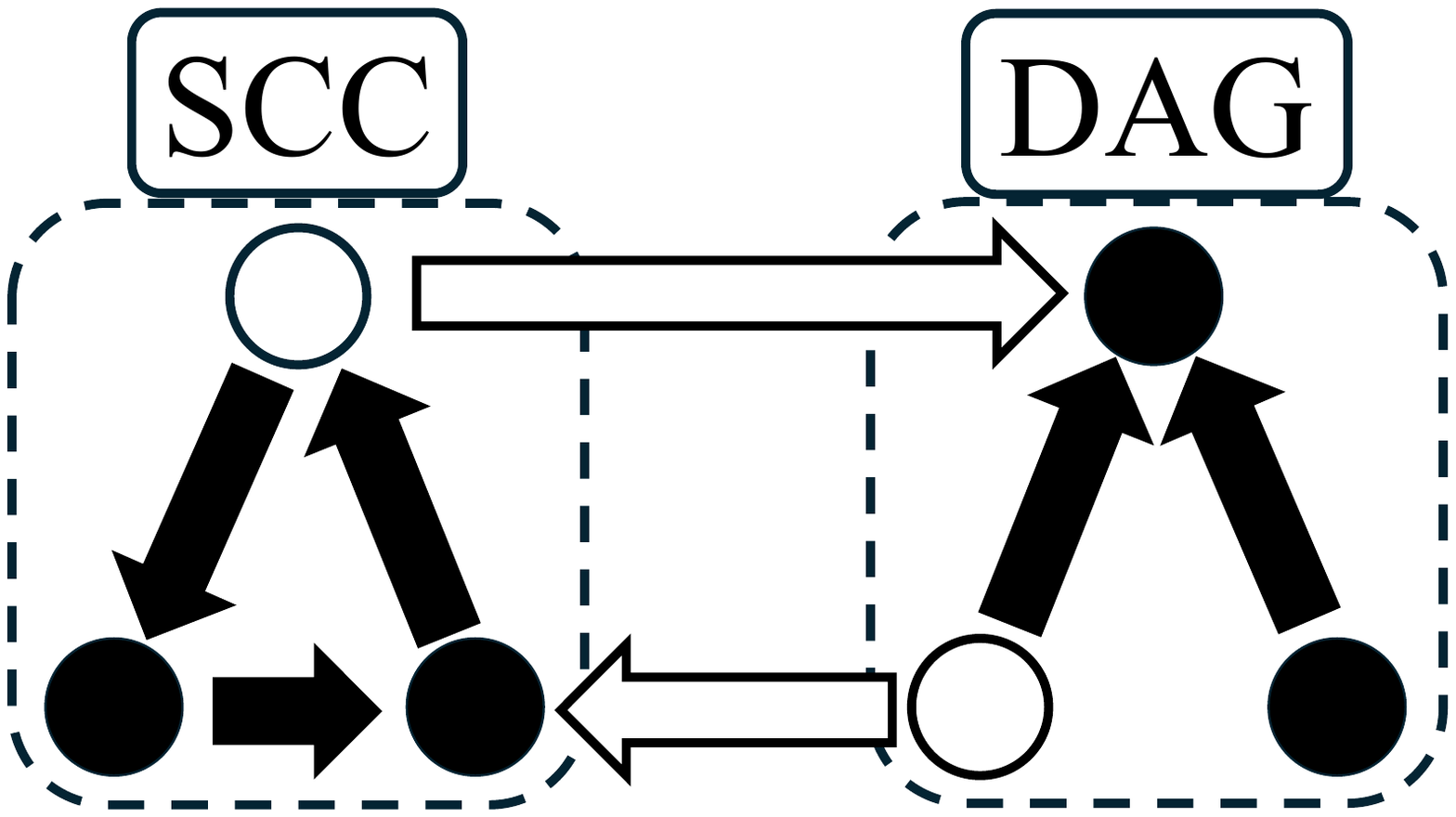}}
    \subfigure[Link through one node ($\looparrowright$).]
        {\includegraphics[width=0.3\textwidth]{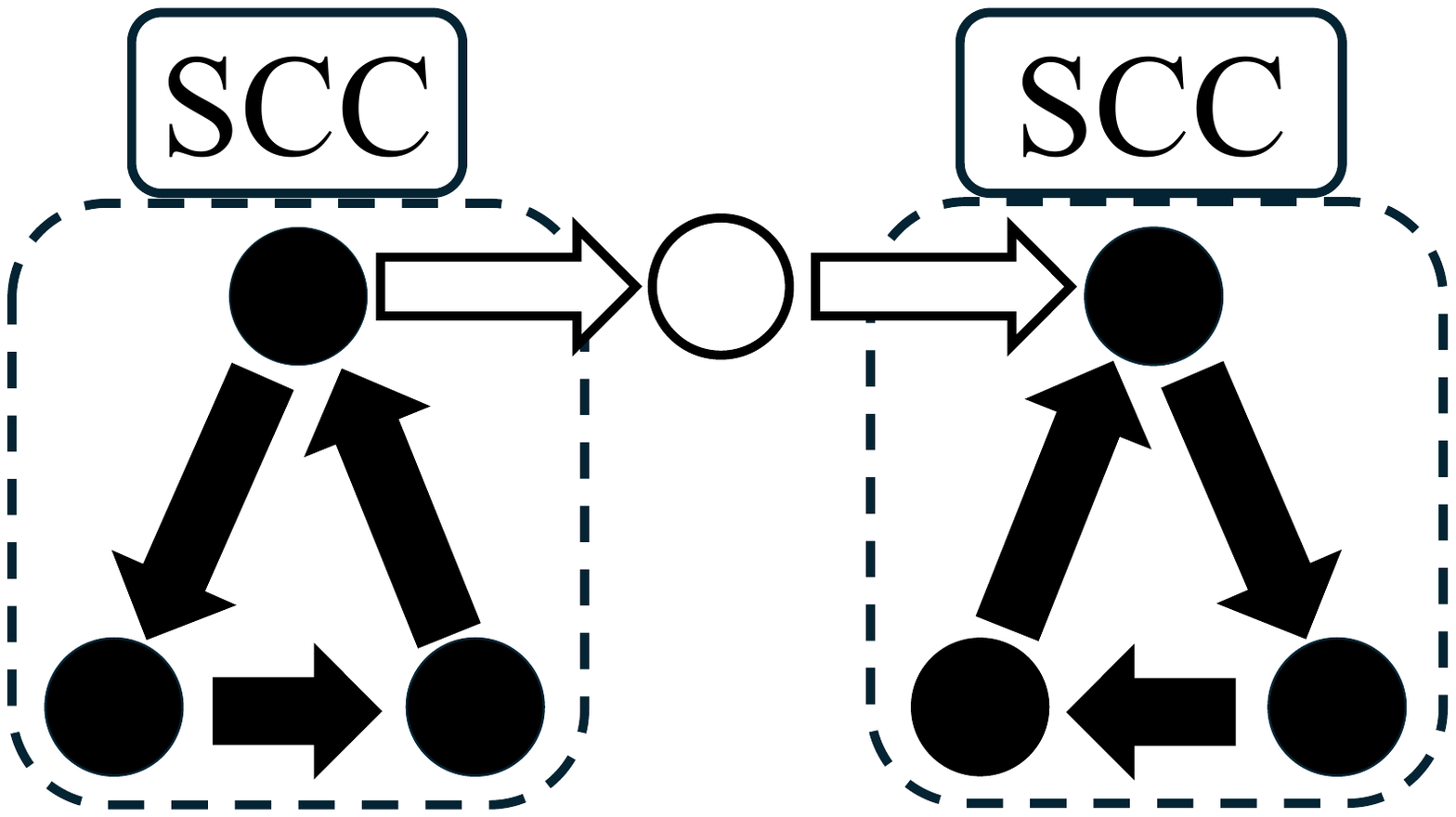}}
    \hspace{0.01\textwidth}
    \caption{Objects and relations for the topological categorisation. In Figures (a), (b), and (c), the considered objects are represented. In Figures (d), (e), and (f), the considered relations are represented.}
    \label{fig:object_relation_illustration}
\end{figure}

In this paper, the inner structure of a directed network is studied as well, but from a complete different perspective. Contrary to the \textit{bow-tie} description, the \textit{core}, \textit{IN-} and \textit{OUT-} components, and relative connections are not considered in this work. Instead, the differentiation between cyclic and acyclic components is the key aspect considered here. A cyclic component is a portion of the network structured as a strongly connected component (SCC), while an acyclic component is a portion of the network structured as a directed acyclic graph (DAG). To understand the difference between cyclic and acyclic components consider the Figures~\ref{fig:object_relation_illustration}(a) and~\ref{fig:object_relation_illustration}(b). In Figure~\ref{fig:object_relation_illustration}(b), the white node in the center of a DAG can move and leave its position, but cannot return to it by following the direction of the arrows. On the other hand, in Figure~\ref{fig:object_relation_illustration}(a), the white node in the center of a SCC can move and leave its position and return to it by following at least 4 different directed \textit{paths} which do not cross the same node twice. These directed paths identify directed simple \textit{cycles}, because they start and end on the same white node's position. For this reason, in this paper, cyclic component and strongly connected component (SCC) are used as synonyms. Similarly, acyclic component and directed acyclic graph (DAG) are used as synonyms as well. Since only cyclic and acyclic components are identified, their connectivity with the rest of the network characterises the categorisation procedure adopted in this paper. In Table \ref{tab:topological_groups_def}, the directed network is split into 14 different components: 11 including nodes and edges, 3 including only edges (see Table~\ref{tab:topological_groups_def} for definitions). This is a comprehensive categorisation which uniquely assign each node and edge into one and only one of those categories, and therefore, one and only one related network component. 

\begin{figure}[!htbp]
    \centering
    \subfigure[]
        {\includegraphics[width=0.47\textwidth]{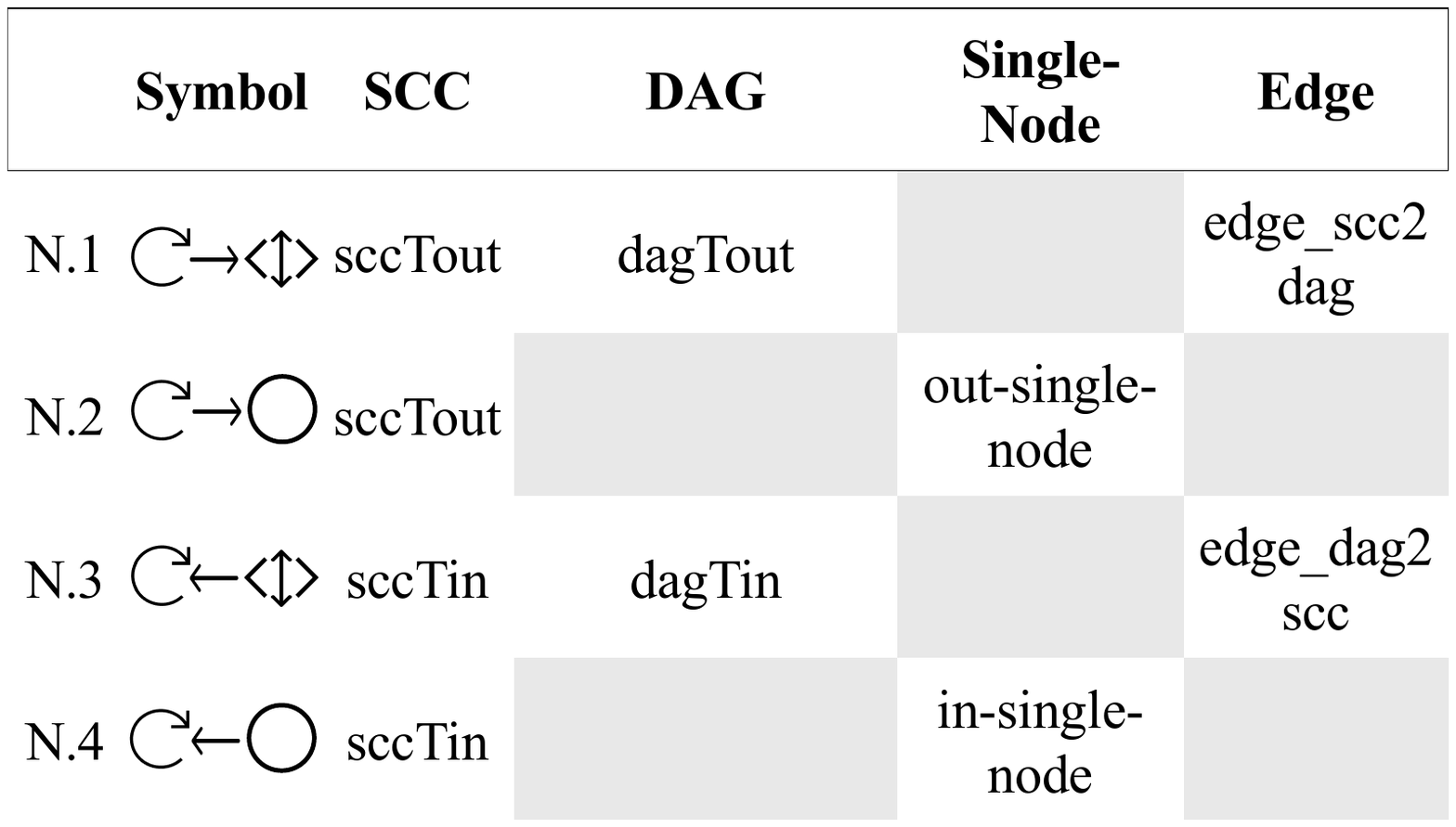}}
    \hspace{0.01\textwidth}
    \subfigure[]
        {\includegraphics[width=0.47\textwidth]{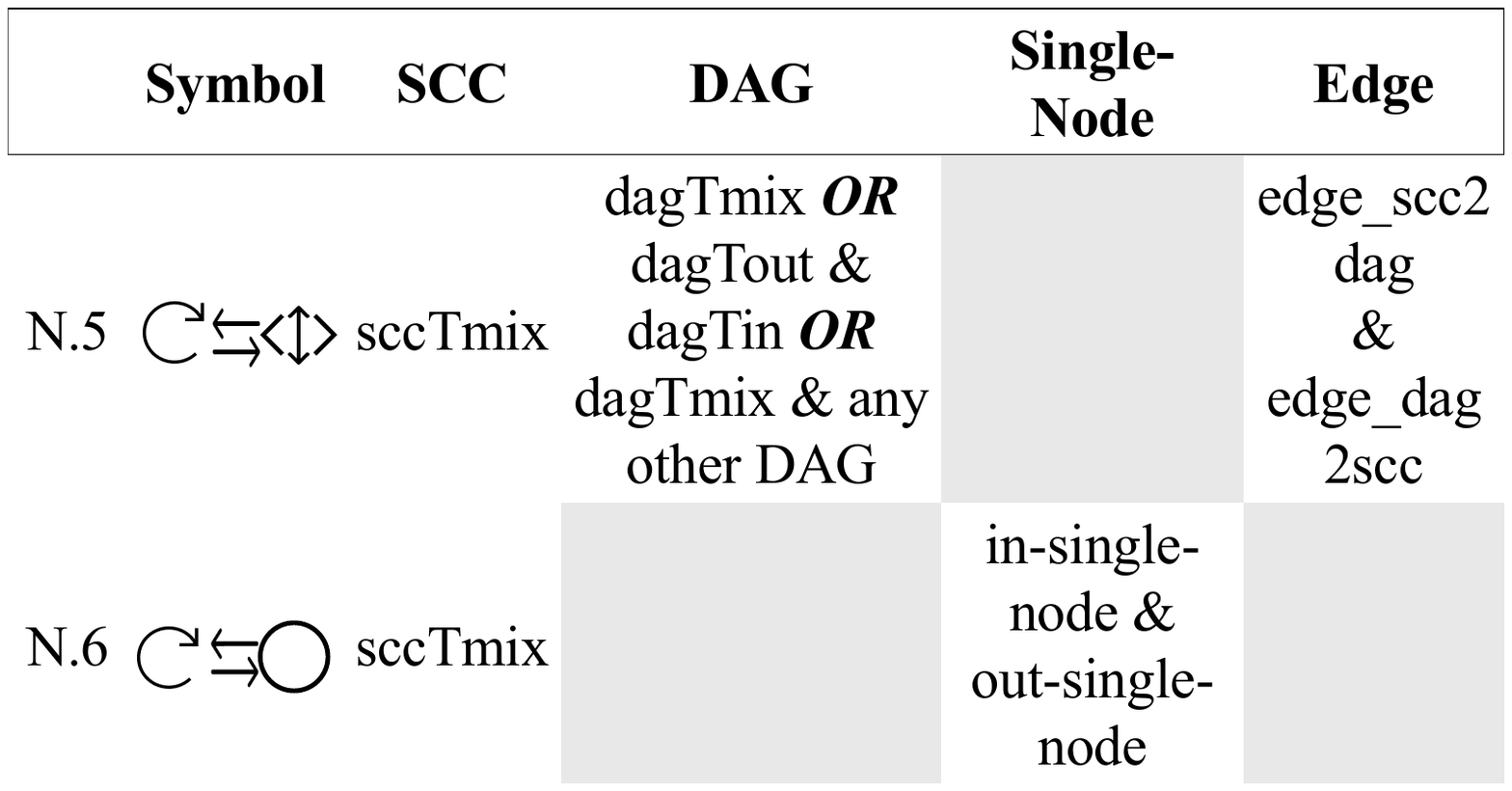}}
    \\
    \subfigure[]
        {\includegraphics[width=0.47\textwidth]{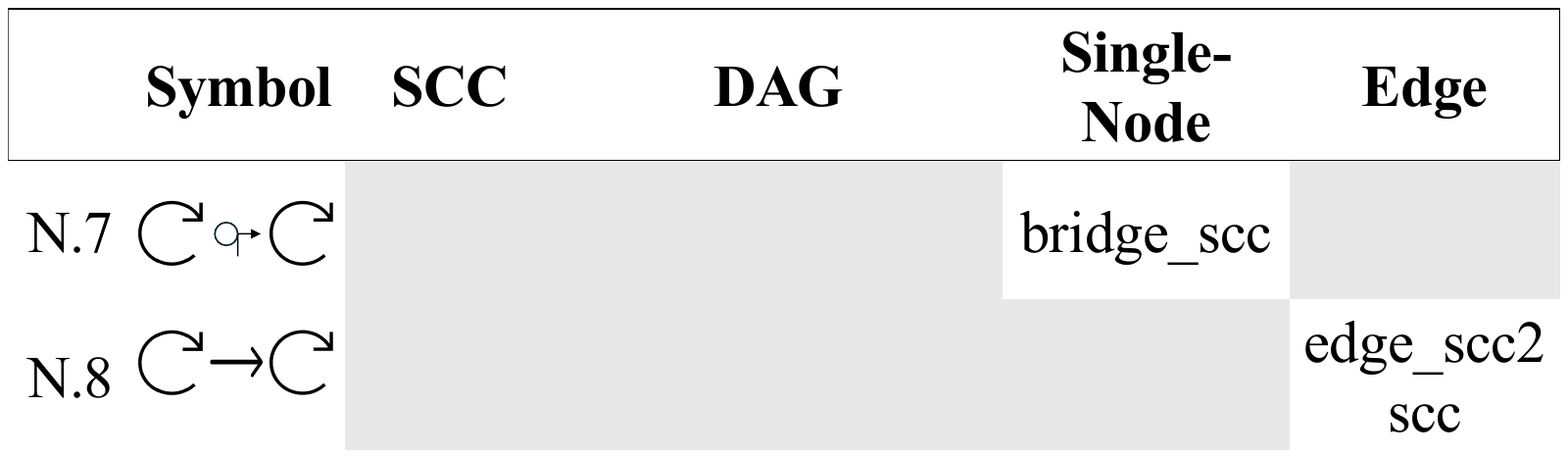}}
    \hspace{0.01\textwidth}
    \subfigure[Legend]
        {\includegraphics[width=0.47\textwidth]{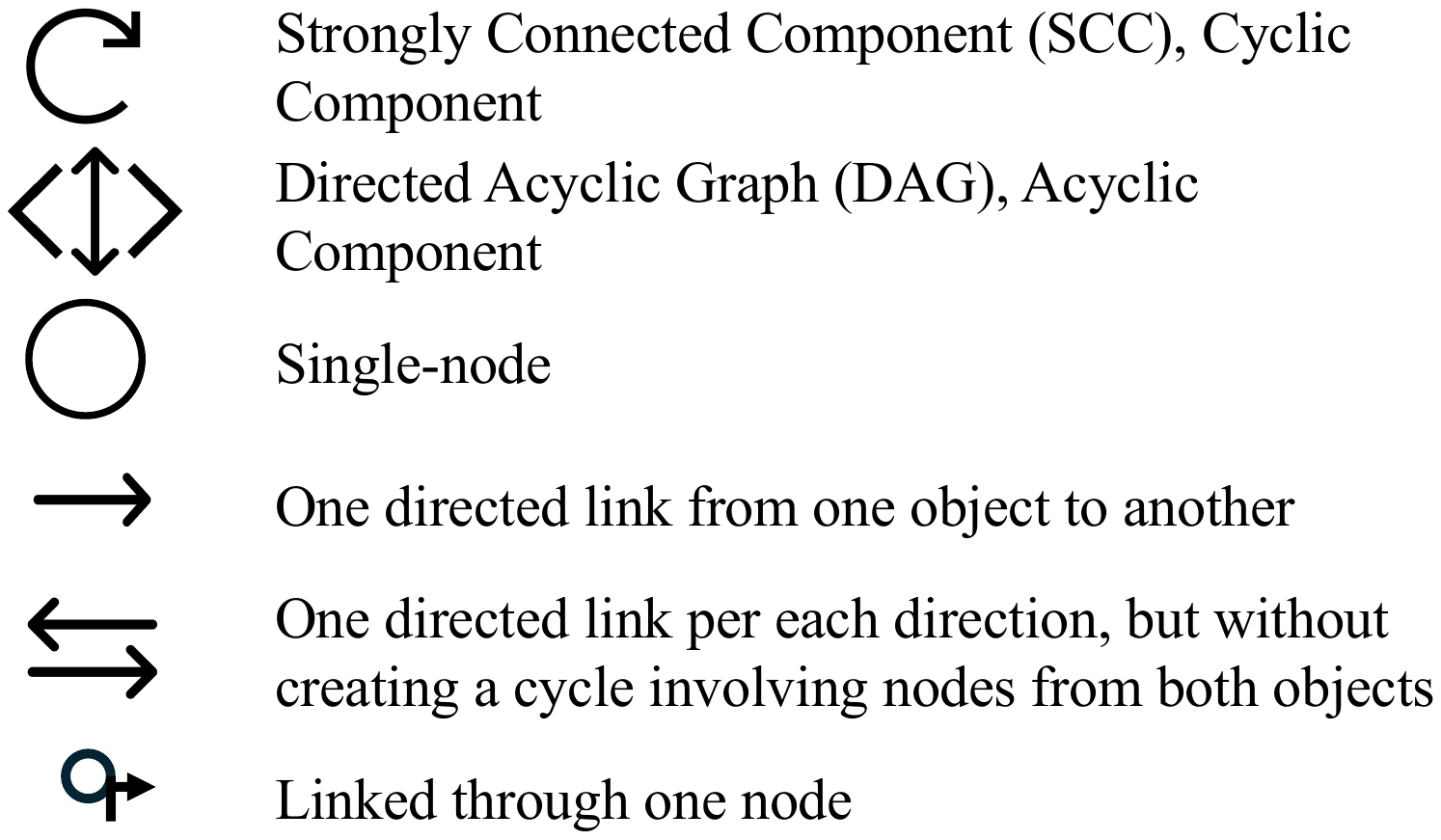}}
    \caption{Label per each topological component. The logic of this categorisation procedure is explained by showing 8 cases in Figures (a)-(c). In each case, there are only two objects in relation to each other. The objects can be either a cyclic component (SCC, $\circlearrowright$), an acyclic component (DAG, $\langle\updownarrow\rangle$), or a single-node ($\bigcirc$). Each pair of objects can have three types of relation exclusively. The first type of relation is a directed link ($\longrightarrow$) from one object to another. The second type of relation ($\leftrightarrows$) is a connection which includes links in opposite direction between two objects (without creating a cycle). The third type of relation is a connection between two objects which involves one node only, which is receiving from one component and sending to another one ($\looparrowright$). In reality, there is often a combination of these 8 cases represented above. For example, a strongly connected component can receive from a DAG and send to a single-node, and therefore, being identified as a \textit{sccTmix}. Finally, consider also that a \textit{bridge\_scc} is a node receiving from a SCC and sending to another SCC, a behaviour which is described by the relation $ \looparrowright$. As already mentioned, it is important to point out that a single-node sending to or receiving from another node in a DAG is part of that DAG itself, and one single-node sending or receiving from another single-node is a pair which constitutes a DAG.}
    \label{fig:top_cat_illustration}
\end{figure}

The logic of this categorisation is explained in Figure~\ref{fig:object_relation_illustration} and Figure~\ref{fig:top_cat_illustration}. In Figure~\ref{fig:object_relation_illustration}, the objects and relations among them are defined. The objects can either be a cyclic component (SCC, $\circlearrowright$)(Figure~\ref{fig:object_relation_illustration}(a)), an acyclic component (DAG, $\langle\updownarrow\rangle$)(Figure~\ref{fig:object_relation_illustration}(b)), or a single-node ($\bigcirc$)(Figure~\ref{fig:object_relation_illustration}(c)). Every pair of objects can have three types of relation between each other. Please note that each relation considered here between each pair of objects cannot change the nature of the object itself. For instance, a cyclic component cannot becomes an acyclic component by adding any of those considered relations between them. The first type of relation considered here is a directed link ($\longrightarrow$) from one object to another (Figure~\ref{fig:object_relation_illustration}(d)). A directed link corresponds to one or more transactions (flow of currency) from one object to another following the direction of that link. The second type of relation ($\leftrightarrows$) is a connection between two objects which includes links in opposite direction, but without creating a cycle which involves nodes from both objects. This type of relation is illustrated in Figure~\ref{fig:object_relation_illustration}(e), where we could imagine that white nodes can move and jump into the opposite component following their white arrows, but we also notice that they eventually cannot return back to their original position by following the direction of the arrows. Obviously, this also implies that the any two nodes from two different components cannot exchange back and forth, otherwise they would create a cycle of length 2. Finally, a third type of relation is a connection between two components which involves one node only ($\looparrowright$). In other words, one node is receiving from one component and sending to another component (see Figure~\ref{fig:object_relation_illustration}(f)). For this last case, please note the difference between Figure~\ref{fig:object_relation_illustration}(c), Figure~\ref{fig:object_relation_illustration}(d), and Figure~\ref{fig:object_relation_illustration}(f). In Figure~\ref{fig:object_relation_illustration}(d), one node connects a SCC to a DAG (i.e. a chain of nodes is part of the DAG); in Figure~\ref{fig:object_relation_illustration}(f) one node receives from one SCC and sends to another; finally, in Figure~\ref{fig:object_relation_illustration}(c) a node is either sending or receiving from a SCC. Obviously, a single-node sending to or receiving from another node in a DAG is part of that DAG itself. Similarly, one single-node sending or receiving from another single-node is a pair which constitutes a DAG. The categorisation procedure is further explained in Figure~\ref{fig:top_cat_illustration}.

In the previous paragraph, objects and relations for the topological categorisation were introduced. In this paragraph, each object is categorised based on its relation with another object. In Figure~\ref{fig:top_cat_illustration}, the logic of this categorisation is explained by showing 8 simple cases, where there are only two objects in relation to each other. Firstly, four types of cyclic components are defined: \textit{sccTin}, \textit{sccTout}, \textit{sccTmix}, and \textit{scc0}. While \textit{scc0} is simply a strongly connected component only connected to other SCCs or isolated, the other categories can emerge in one of those cases represented in Figures ~\ref{fig:top_cat_illustration}(a) and ~\ref{fig:top_cat_illustration}(b). A \textit{sccTout} component emerges if a SCC sends to a DAG or single-node (N.1 and N.2, Figure~\ref{fig:top_cat_illustration}(a)). A \textit{sccTin} component emerges if a SCC receives from a DAG or single-node (N.3 and N.4, Figure~\ref{fig:top_cat_illustration}(a)). A \textit{sccTmix} component emerges if a SCC sends to and receives from a DAG and/or single-node (N.5 and N.6, Figure~\ref{fig:top_cat_illustration}(b)). 

Secondly, the typology for acyclic components is also categorised as: \textit{dagTin}, \textit{dagTout}, \textit{dagTmix}, and \textit{dag0}. In this case, \textit{dag0} is an isolated acyclic component, while the other categories can also emerge in one of those cases represented in Figure ~\ref{fig:top_cat_illustration}(a) and ~\ref{fig:top_cat_illustration}(b). A \textit{dagTout} component emerges if a DAG receives from a SCC (N.1, Figure~\ref{fig:top_cat_illustration}(a)). A \textit{dagTin} component emerges if a DAG sends to a SCC (N.3, Figure~\ref{fig:top_cat_illustration}(a)). A \textit{dagTmix} component emerges if a DAG sends to and receives from a SCC (N.5, Figure~\ref{fig:top_cat_illustration}(b)), obviously without creating a cycle with it. 

The edges between SCC and DAG are considered as a separate category: from cyclic component to an acyclic component (\textit{edge\_scc2dag} in N.1 and N.5 in Figure~\ref{fig:top_cat_illustration}(a) and ~\ref{fig:top_cat_illustration}(b)) and from an acyclic component to a cyclic component (\textit{edge\_dag2scc} in in N.3 and N.5 in Figure~\ref{fig:top_cat_illustration}(a) and ~\ref{fig:top_cat_illustration}(b)). Similarly, edges between different SCCs (\textit{edge\_scc2scc}) are also considered separately (N.8 in Figure~\ref{fig:top_cat_illustration}(c)). 

The only special case left to be discussed are single-nodes, which are nodes that cannot really be associated neither to cyclic nor acyclic components. Splitting a directed network only into cyclic and acyclic components (and link between them) is not be sufficient to identify a comprehensive and unique categorisation for each node and edge in the network. Indeed, the inclusion of single-nodes should complete its description. A single-node receiving from one SCC and sending to another SCC is called \textit{bridge\_scc} (N.7 in Figure~\ref{fig:top_cat_illustration} and in Figure~\ref{fig:object_relation_illustration}(f)). A single-node only sending to SCCs is called \textit{in-single-node} (N.4 in Figure~\ref{fig:top_cat_illustration} and Figure~\ref{fig:object_relation_illustration}(c)). A single-node only receiving from SCCs is called \textit{out-single-node} (N.2 in Figure~\ref{fig:top_cat_illustration} and Figure~\ref{fig:object_relation_illustration}(c)). Any connection to a DAG will categorise the single-node as a part of that DAG itself. A single-node connected to another single-node will obviously create a DAG. In Table \ref{tab:topological_groups_numb}, an analytical description of these categories is reported. Since each category corresponds to a uniquely identified network component, the sum of the volume of each component corresponds to the total volume of the network. For this reason, we conclude that this topological categorisation completely and successfully describe the whole network under examination. Nevertheless, we do not exclude that future studies may find a better way to improve this technique and adapt it to different contexts and purpose.

\begin{table}[!htbp]
    \centering
    \begin{tabularx}{0.9\textwidth}{|X|p{0.11\textwidth}|p{0.52\textwidth}|}
    \hline
        \textbf{Node} &  \textbf{Edge} & \textbf{Definition} \\
         \hline
         sccTmix & = & SCC sending to and receiving from DAG or single-node\\
         in-single-node & = & Single-node sending to SCCs\\
         dagTin & = & DAG sending to SCCs\\
         dag0 & = & Isolated DAG\\
         out-single-node & = & Single-node receiving from SCCs\\
         scc0 & = & Isolated SCC, or connected to other SCCs through a \textit{edge\_scc2scc} or a \textit{bridge\_scc}\\
         sccTin & = & SCC receiving from DAG or single-node\\
         dagTmix & = & DAG sending to and receiving from with SCCs\\
         dagTout & = & DAG receiving from SCCs\\
         sccTout & = & SCC sending to DAG or single-node\\
         bridge\_scc & = & Single-node connecting two or more SCCs\\
         -- & edge\_dag2scc & Link from a DAG to a SCC \\
         -- & edge\_scc2dag & Link from a SCC to a DAG \\
         -- & edge\_scc2scc & Link connecting two SCCs\\
         \hline
    \end{tabularx}
    \caption{Description of topological categories for edges and nodes. \textit{SCC} is used as abbreviation for strongly connected component. \textit{DAG} is used as abbreviation of directed acyclic component.
    Note that \textit{edge\_dag2scc}, \textit{edge\_scc2dag}, and \textit{edge\_scc2scc} do not have a corresponding node because their nodes are already assigned to different components. For example, one \textit{edge\_scc2dag} is made of one sender in a \textit{SCC} and one receiver in a DAG. Similarly, \textit{bridge\_scc} is only one single-node receiving from one SCC and sending to another SCC. However, if there is a chain of nodes where the first node is receiving from a SCC and sending to another SCC through its last node, this is considered as \textit{dagTmix}.}
    \label{tab:topological_groups_def}
\end{table}

\subsection{Recirculation} \label{sec:methods:recirculation}

\begin{figure}[!ht]
    \centering
    \includegraphics[width=0.5\textwidth]{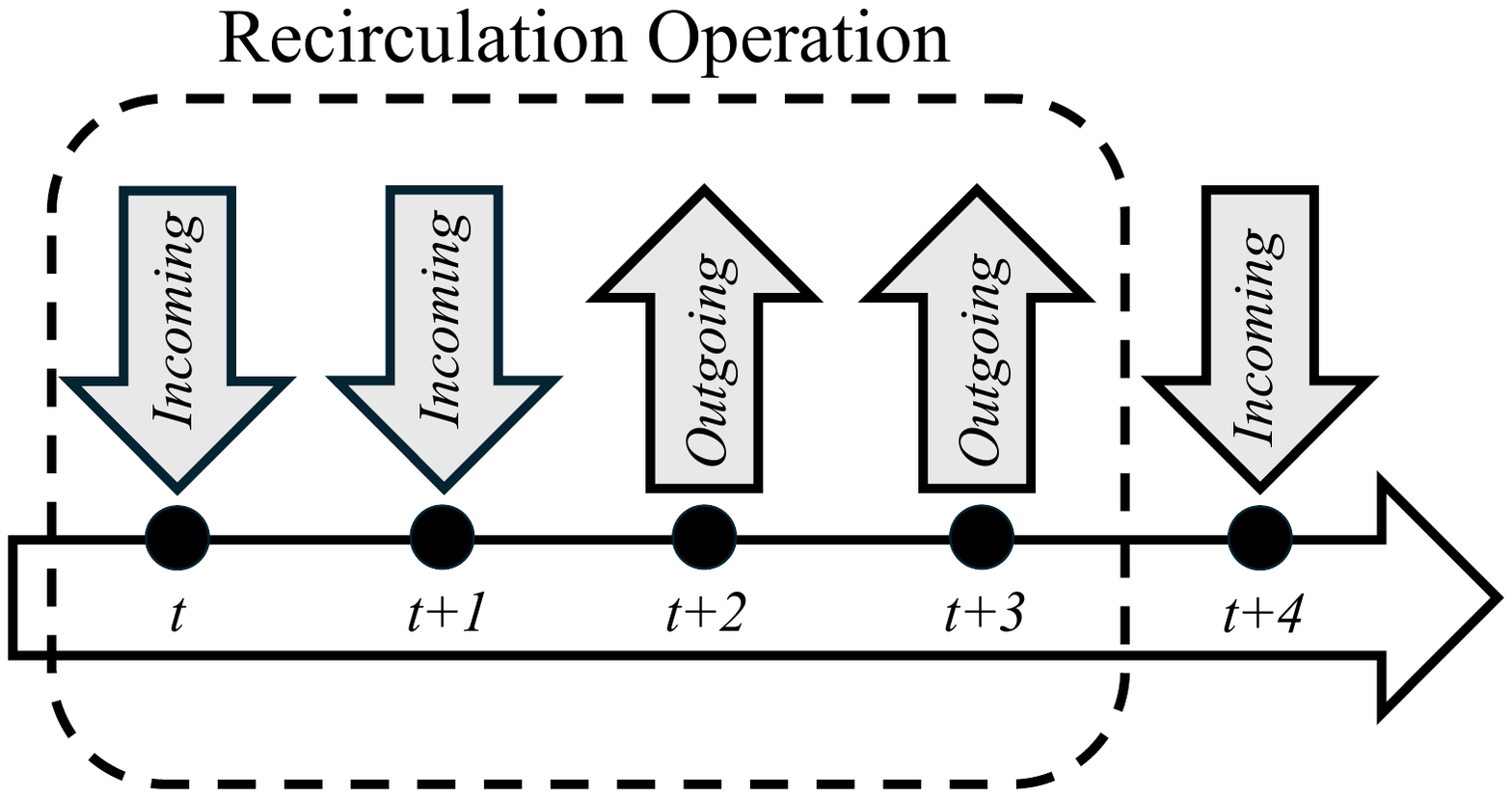}
    \caption{Illustration of a recirculation operation as defined in this paper. One recirculation operation includes many incoming and outgoing operations. The node in the figure receives two incoming transactions at time \textit{t} and \textit{t+1}. After that, it sends currency twice at \textit{t+2} and \textit{t+3}. At \textit{t+4}, it receives again some currency, so the recirculation operation closes.}
    \label{fig:recirculation_illustration}
\end{figure}

In this paper the characterisation of the dynamics of the network is carried out by using the recirculation time. The transfer velocity of circulation was used in a recent work to describe the circulation of Sarafu using this data~\cite{mattssonVelocity}. In that work, Authors were interested in the time between one incoming operation and the first outgoing operation ("first-in first-out"~\cite{mattsson_trajectories_2021}), and then averaging this quantity to define the "holding time" per each user. The Authors then analyse the "holding time" in relation to business sectors and geographic areas. As described in Figure \ref{fig:recirculation_illustration}, the recirculation time is here measured as the time difference between the first of all the incoming operations and the last of all the outgoing operations before another incoming operation arrives. Moreover, the recirculation time is not aggregated (or averaged) per each user. Indeed, each individual is assigned to one or more temporal categories, according to the speed of their recirculation operations (Section \ref{sec:results:recirculation}). Instead of focusing on circulation in business sectors and geographic areas, the recirculation time is then analysed in relation to these predefined topological categories.

The frequency (or speed) of recirculation is therefore defined as the time difference between the first incoming operation (at time \textit{t} in Figure~\ref{fig:recirculation_illustration}) and the last outgoing operation (at time \textit{t+3} in Figure~\ref{fig:recirculation_illustration}). In the example of Figure~\ref{fig:recirculation_illustration} the frequency (or speed) is equal to 2. This value is then used to categorise recirculating operations from low to high frequency. The distribution of recirculation frequencies is left-skewed, because most of the recirculating operations (or recirculations) are happening in a very short time. The operations are therefore split into four sets based on the quartiles of the frequency distribution (see Table \ref{tab:rtmax_tab}). For instance, the first quartile of \textit{19 minutes, 39 seconds} defines the temporal category \textit{HFQ1}; its most frequent value (or \textit{Mode}) is \textit{104 seconds} and this is also the mode of the whole recirculation time distribution. About half of those operations is happening within about 10 hours (\textit{HFQ2}) and 75\% of them within less than two days (\textit{HFQ3}). It is useful to analyse separately these temporal categories - from "high" to "low" frequencies of recirculation. Following this reasoning, recirculation operations are categorised. Accordingly, users can be assigned to one or more temporal category based on the speed of their recirculation operations. For example, if a user did one recirculation within 5 seconds and then another recirculation operation within 5 days, then the user is categorised as \textit{HFQ1-LFQ3}. In the Section~\ref{sec:results:recirculation}, the result of this type of analysis on users is showed.

\begin{table}[ht]
    \centering
    \begin{tabularx}{\textwidth}{|p{0.1\textwidth}|X|X|X|p{0.1\textwidth}|}
        \hline
        \textbf{Recirc.} & \textbf{From} & \textbf{To} & \textbf{Mode}  &  \\
         \hline
         HFQ1 & $\sim{1}$ second & 19.0 minutes, 39.0 seconds & 104 seconds (158 times) & 0-25\%\\
         \hline
         HFQ2 & 19.0 minutes, 40.0 seconds & 10.0 hours, 3.0 minutes & 28 minutes, 20 seconds (12 times) & 25\%-50\%\\
         \hline
         HFQ3 & 10.0 hours, 3.0 minutes & 1.0 day, 21.0 hours & 22 hours, 30 minutes (6 times) & 50\%-75\%\\
         \hline
         LFQ3 & 1.0 day, 21.0 hours & 50.0 weeks, 6.0 days & 1 day, 22 hours (3 times) & 75\%-100\%\\
         \hline
    \end{tabularx}
    \caption{Split of recirculation time distribution along quartiles. The time of recirculation shows a tendency to recirculate within a day. In fact, 75\% of recirculations are happening before 1 day and 21.0 hours (Q3).}
    \label{tab:rtmax_tab}
\end{table}

\section{Data} \label{sec:data}

The data used in this paper are timestamped transactions from the Sarafu system in Kenya between 25 January 2020 and 15 June 2021~\cite{ruddick_sarafu_2021}. Further user information is available in the data: geographic location, business sector, and gender. In this period, the currency was used as part of a COVID-19 disaster-response intervention in Mukuru kwa Njenga slum, Nairobi and in Kisauni, Mombasa in collaboration with the Kenyan Red Cross~\cite{mattssonSarafu}. The Sarafu token was used as a cash transfer program in local vouchers. This means that each user could spend the Sarafu token only in local businesses. Furthermore, local businesses could re-use the Sarafu token among each other, as far as everyone had joined the network. For a limited time period, some donors backed the initial fund, so that the cash-out in Kenyan Shillings was for that short time limited to users and vendors through savings groups (or \textit{Chamas}), under certain constraints. 

It is also important to mention that since 2017 the Kinango (Kwale county) area has been targeted by Grassroots Economics for specific development interventions: donations have been collected to build community-owned assets with the purpose of enhancing community socio-economic resilience (e.g. maize milling, refrigeration, water storage equipment, etc.)~\cite{mattssonSarafu}. In fact, in the data, Kinango, Mukuru kwa Njenga slum, and Kisauni are the most active geographic areas. Nonetheless, about 86\% of the users are coming from one of those treated areas, so any comparison with untreated areas would be unbalanced.

\textit{Chamas} are informal cooperative groups used to pool and invest savings and typical in East Africa, which play a significant role in the Sarafu network~\cite{mattssonSarafu, mattssonCirculation, barinaga_2020, cooperative_sarafu_2023, temporal_sarafu_2022}. In the data, they are identifiable as \textit{group accounts} (formal savings group) or \textit{savings} business accounts (informal savings groups), but they are excluded in the analyses carried out in this paper. In fact, this paper is only focused on user accounts because the main aim is to identify individual strategy in engaging with the Sarafu economic network.

The main focus of this paper is to study the economic behaviour of Sarafu users. For this reason, the financial network of savings groups (or \textit{Chamas}) was removed from the dataset. The resulting transaction network contains 39,433 users, 360,117 transactions, and a total volume of 182,605,612.73 Sarafu (in parity with Kenyan Shillings). In this network of individual users 28,709,009.6 Sarafu were disbursed mostly to new members and to reward existing members (Section \ref{sec:discussion} for details). The users are grouped into 10 main geographic areas: Kilifi, Kinango (Kwale County), Kisauni (Mombasa County), Mombasa, Nairobi, Rural Counties, Mukuru (Nairobi County), Nyanza County, Turkana County, and other/unknown. See Supplementary Material (Document A, Section "Data") for further information on geographic areas and business sectors.

As described in a previous work~\cite{mattssonCirculation}, the degree distributions of this network are heavy-tailed. The degree distributions are built by aggregating all the transactions happening between each pair of nodes and preserving their directionality. The in-degree and out-degree distributions can be well approximated by power-laws, respectively with exponents 1.53 and 1.47 (see Supplementary Material, Document A, Section "Data"). Similarly, the distribution of the number of transactions per link also behaves as a power-law with an exponent 1.44. This means that a very high number of links have one or few transactions happening on them, while a few links are responsible for a very high number of transactions. The distribution of volume per link is also well approximated by a power-law distribution with exponent 1.85. The Pearson correlation between the distribution of number of transactions and total weight per edge is equal to 0.58 significant at 1\% (p-value<0.01).

\section{Results} \label{sec:results}

\subsection{Network Topology} \label{sec:results:top_cat}

\begin{figure}[!htbp]
  \centering
  \subfigure[Subgraph]{\includegraphics[width=5in]{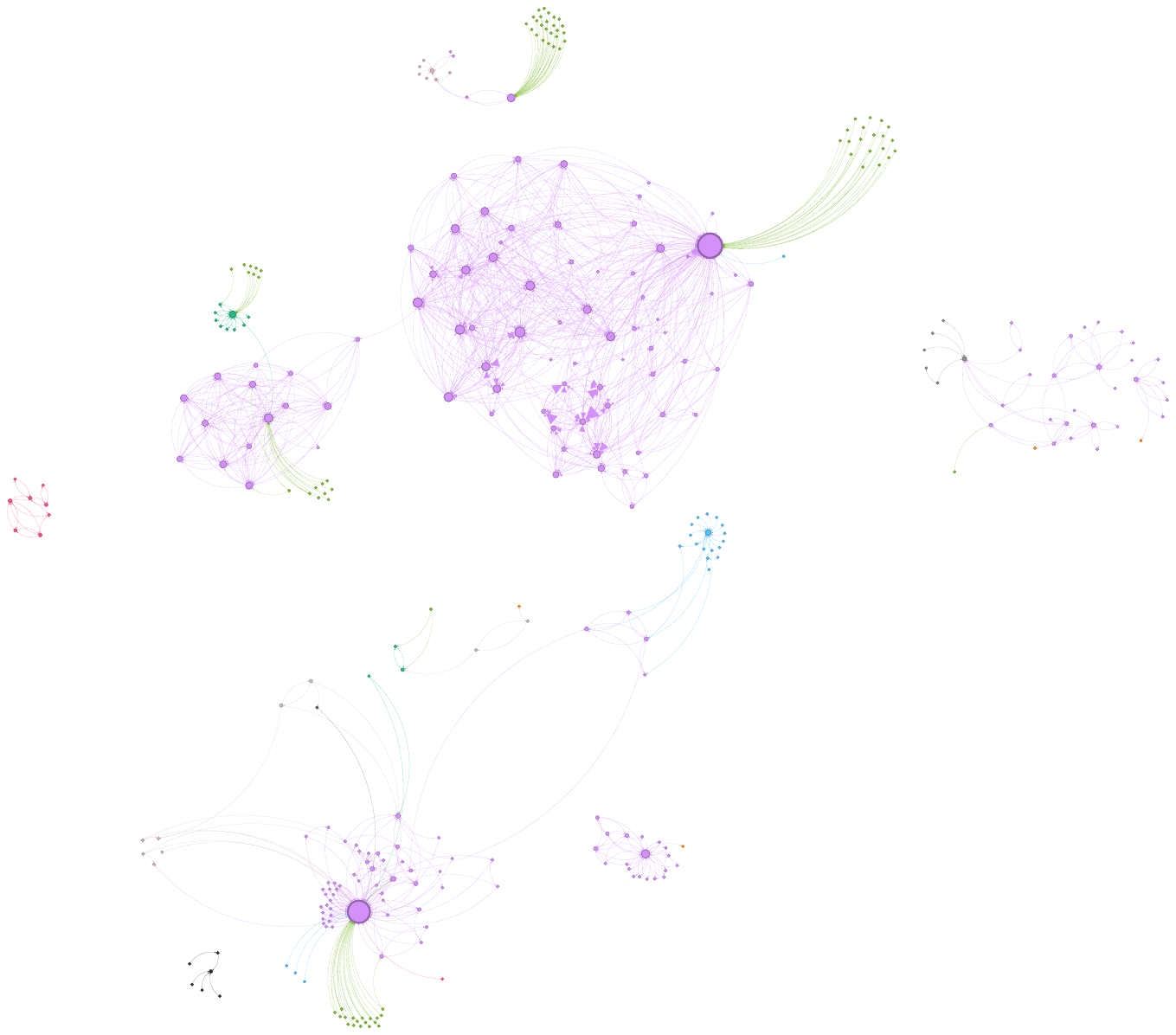}}
  \hspace{0.1\textwidth} 
  \subfigure[Legend]{\includegraphics[width=0.2\textwidth]{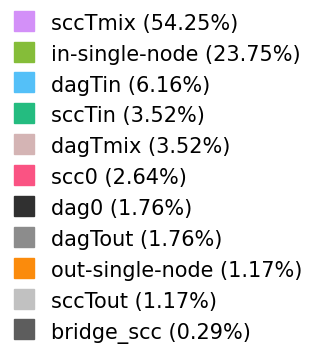}}
  \caption{Subgraph of 341 nodes and 1060 directed links. The subgraph is created by merging ego graphs at depth 1, 2, and 3 of 11 nodes, one random node per each topological category. The \textit{in-single-nodes} (in moderate green) are all connected to a few hubs of \textit{sccTmix} components (in very soft violet). Similarly, one central node in a \textit{dagTin} component (in soft blue) is collecting from other nodes and then sending to other not central \textit{sccTmix} nodes. Furthermore, \textit{dagTmix} (in grayish red, bottom left) and \textit{dagTout} (in dark gray, top right) can also be observed attached to two different \textit{sccTmix} components. The \textit{dagTout} is made of one hub receiving from other (\textit{sccTmix} and \textit{dagTout}) nodes, while \textit{dagTmix} is made of dyads. Finally, on the left side we can also see isolated \textit{dag0} (on the bottom, in very dark gray, almost black) and \textit{scc0} (on the top, in soft pink) components. This plot is made out of a sample of ego graphs, hence, it is very likely that some connections are missing within the same components. In the legend, the proportion of users per each category in parenthesis.}
\label{fig:topology_subgraph}
\end{figure}

In this section, the results of the topological analysis are reported and the empirical values are compared to null models to test their statistical significance. A strongly connected component (\textit{SCC}) is a subgraph of a directed network where any node is involved in one or more directed simple cycles. As consequence, in a transaction network every node in a \textit{SCC} is both buying and selling. In a transaction network the importance for a node to be involved in cycles has been empirically explored in a few recent works~\cite{iosifidis, mattssonCirculation, fleischman_liquidity-saving_2020, fleischman_balancing_2020}. In this work, the topological categorisation is used to distinguish between cyclic components (i.e. \textit{SCC}s), acyclic components (i.e. \textit{DAGs}) and single-nodes. Cyclic components are identified by strongly connected components and so the prefix \textit{scc-} is used. Acyclic components are identified by directed acyclic graphs and so the prefix \textit{dag-} is used. The complete topological categorisation used in this paper is explained in Section~\ref{sec:methods:topology}. In Table \ref{tab:topological_groups_def}, a short definition per each category is reported. In Figure \ref{fig:topology_subgraph}, a randomly sampled subgraph illustrates the network of users colored by their topological category. The Figure \ref{fig:topology_subgraph} is a quick representation of the main findings discussed in this section.

\begin{figure}[!ht]
    \centering
    \subfigure[DAGs]{\includegraphics[width=0.45\textwidth]{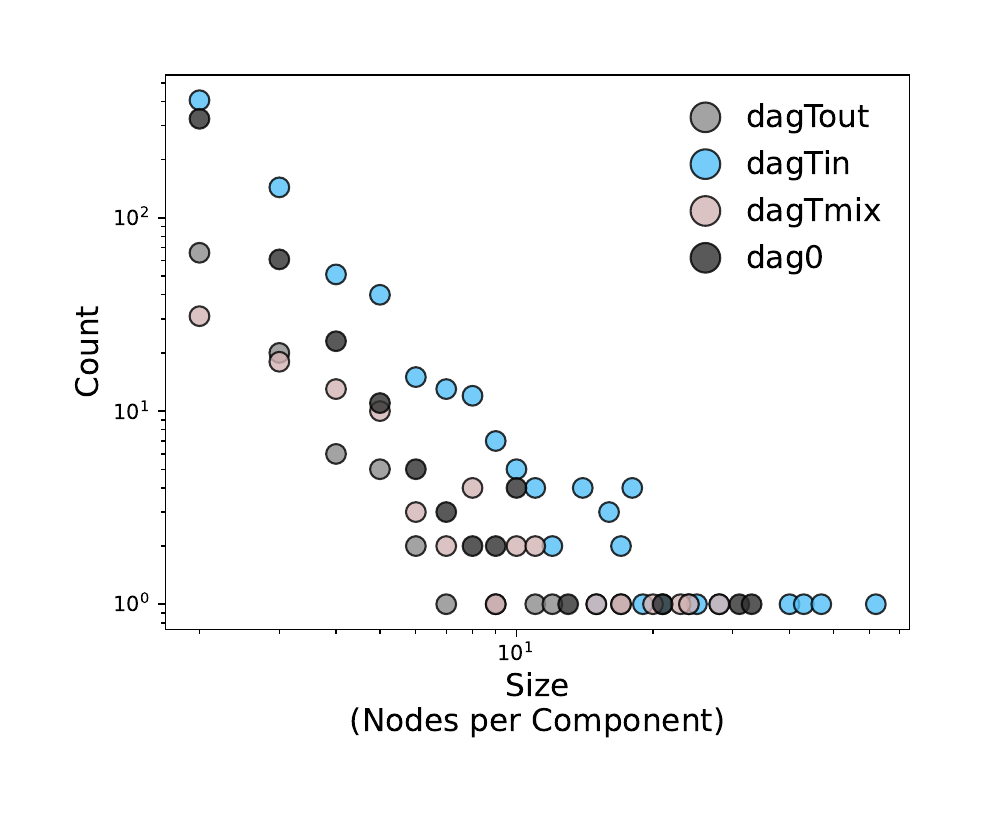}}
    \hspace{0.01\textwidth} 
    \subfigure[SCCs]
    {\includegraphics[width=0.45\textwidth]{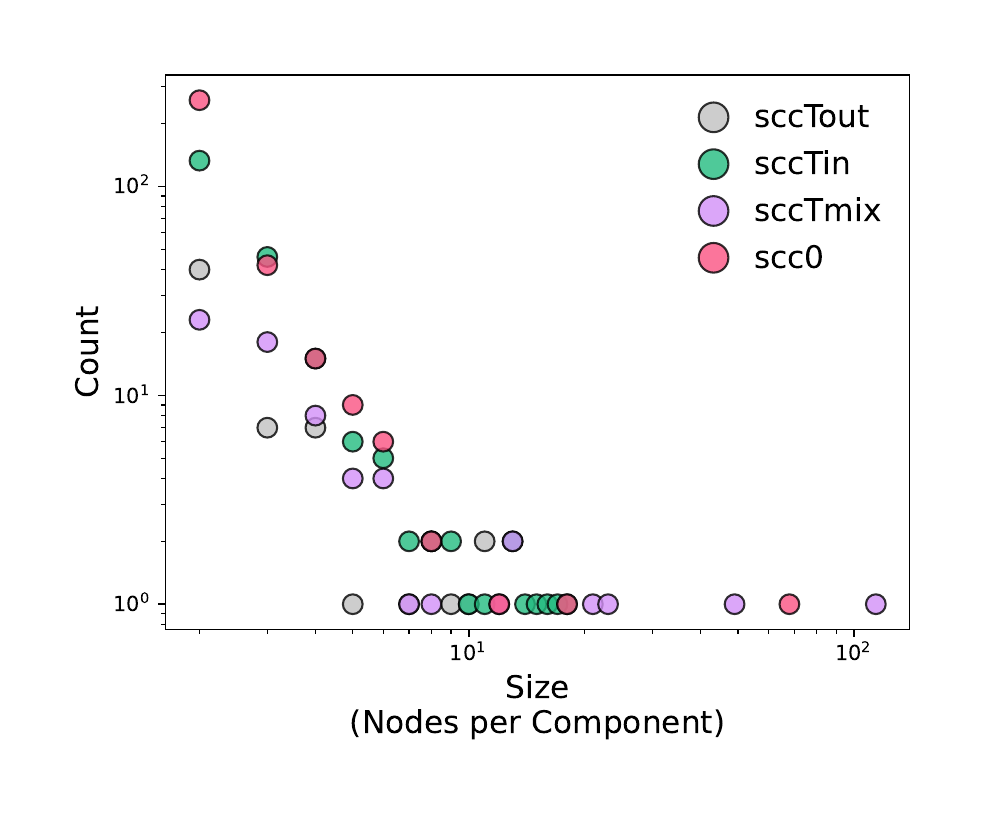}}
    \caption{Size of SCCs and DAGs as weakly connected components, in terms of number of nodes. The largest strongly connected component (LSCC) is excluded from this plot, which has 19,737 nodes, 315,602 transactions, and a volume of 174,919,583.45 Sarafu. After removing the LSCC, the network is left with 19,696 nodes and 15,663 transactions, and a volume of 2,511,366 Sarafu. The weakly connected components are detected after removing the LSCC. Note that the sum of LSCC and not-LSCC is not equal to the total sum of whole network. Since weakly connected components within to the two are considered, the connections between them are excluded from the calculation.}
    \label{fig:scc_dag_size}
\end{figure}

Using the topological categorisation, the network is split into several components, as described above. In Figure~\ref{fig:scc_dag_size}, the size of SCC and DAG detected (as weakly connected components, WCCs) is calculated after removing the largest strongly connected component from the network. The largest strongly connected component has 19,737 nodes, 315,602 transactions, and a volume of 174,919,583.45 Sarafu. The size of DAGs ranges from 2 to 62 nodes, while the size of SCC ranges from 2 to 114 nodes. In Table \ref{tab:topological_groups_numb}, network features per each type of component is reported. The biggest category is \textit{sccTmix} which includes also the the largest strongly connected component. The second largest category is the one of \textit{in-single-node} nodes, namely nodes that are sending currency to another node in a \textit{SCC}(see Figure \ref{fig:topology_subgraph}, as an example).

\begin{table}[!htbp]
\centering
\begin{tabularx}{0.85\textwidth}{|l|l|p{0.04\textwidth}|p{0.05\textwidth}|p{0.06\textwidth}|p{0.07\textwidth}|p{0.07\textwidth}|l|}
    \hline
   \textbf{Node Label} & \textbf{Edge Label} & \textbf{SCCs} & \textbf{WCCs} & \textbf{Nodes} & \textbf{Dir.Links} & \textbf{TX} & \textbf{Volume}\\ 
   \hline
   
    sccTmix & =    & 67           & 67           & 20,173 & 107,067 & 318,567 & 175,590,852.21 \\ \hline
    in-single-node & =  & --           & 1,520         & 14,889 & 15,613  & 24,062  & 4,203,649.49  \\ \hline
    dagTin & =       & --             & 723          & 2,642  & 1,947   & 2,300   & 411,191.95 \\ \hline
    dag0 & =         & --             & 441          & 1,205  & 766    & 906    & 124,367.0 \\ \hline
    out-single-node & = & --             & 534          & 1,749  & 1,224   & 1,620   & 98,822.03 \\ \hline
    scc0 & = & 336          & 336          & 899   & 1,159   & 1,825   & 161,622.65 \\ \hline
    sccTin & =      & 220          & 220          & 699   & 1,165   & 3,688   & 634,183.02 \\ \hline
    dagTmix & =      & --             & 92           & 460   & 389    & 502    & 99,747.96 \\ \hline
    dagTout & =      & --             & 104          & 309   & 207    & 280    & 30,067.0 \\ \hline
    sccTout & =  & 62           & 62           & 198   & 350    & 937    & 49,513.0 \\ \hline
    bridge\_scc & =  & --             & 20           & 75    & 57     & 78     & 21,280.0 \\ \hline
    -- & edge\_dag2scc      & --             & 415          & 2,248  & 2,008   & 3,272   & 709,891.184 \\ \hline   
    --& edge\_scc2dag      & --             & 178          & 622   & 490    & 777    & 126,142.14 \\ \hline
    --& edge\_scc2scc      & --             & 256          & 1,112  & 894    & 1,303   & 344,283.09 \\ \hline
\end{tabularx}

\caption{Size for each topological category.}
\label{tab:topological_groups_numb}
\end{table} 

\begin{figure}[!htbp]
  \centering
  \subfigure[Statistical significance for the \textbf{sccTmix} category.]{\includegraphics[width=0.4\textwidth]{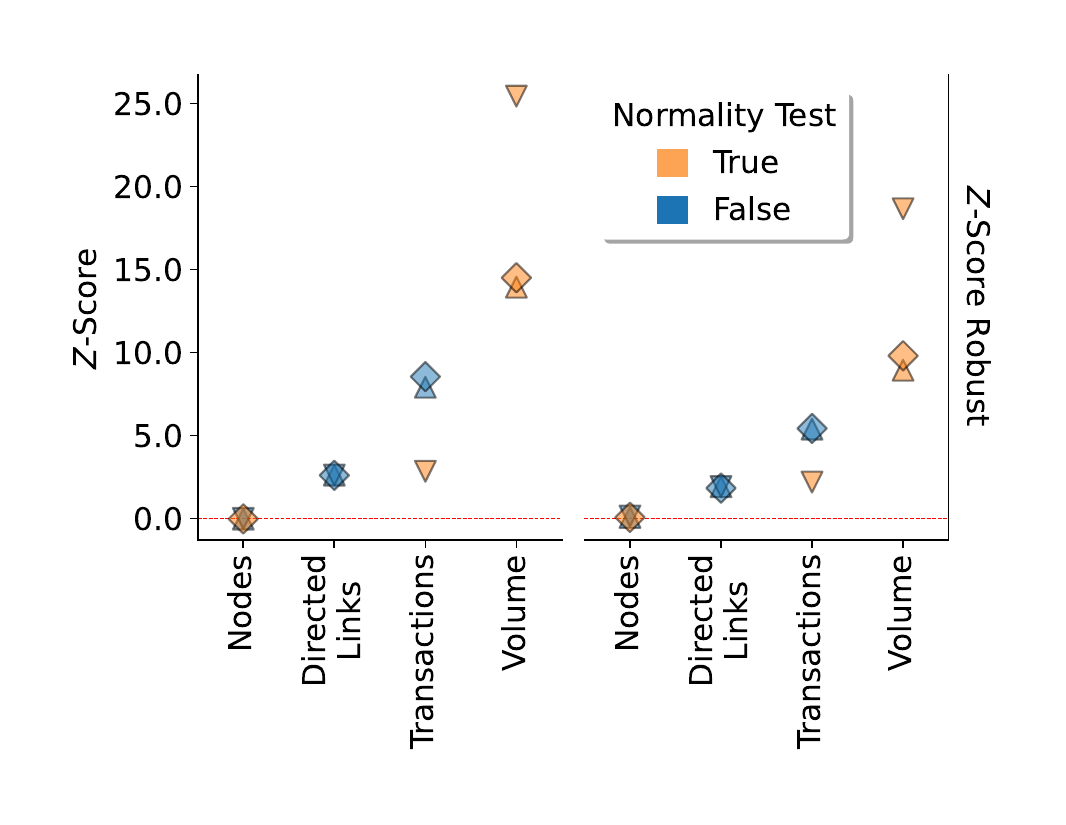}}
  \\ 
  \subfigure[Statistical significance for the \textbf{in-single-node} category.]{\includegraphics[width=0.4\textwidth]{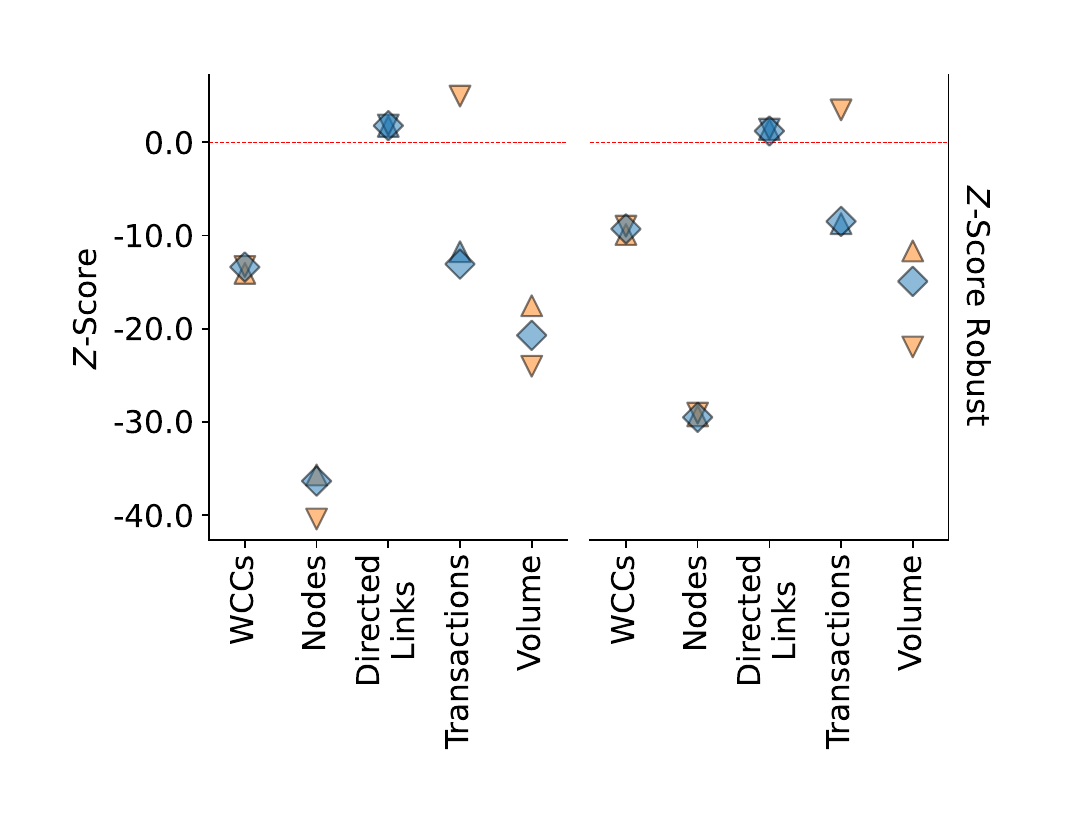}}
  \hspace{0.02\textwidth}
  \subfigure[Statistical significance for the \textbf{out-single-node} category. ]{\includegraphics[width=0.4\textwidth]{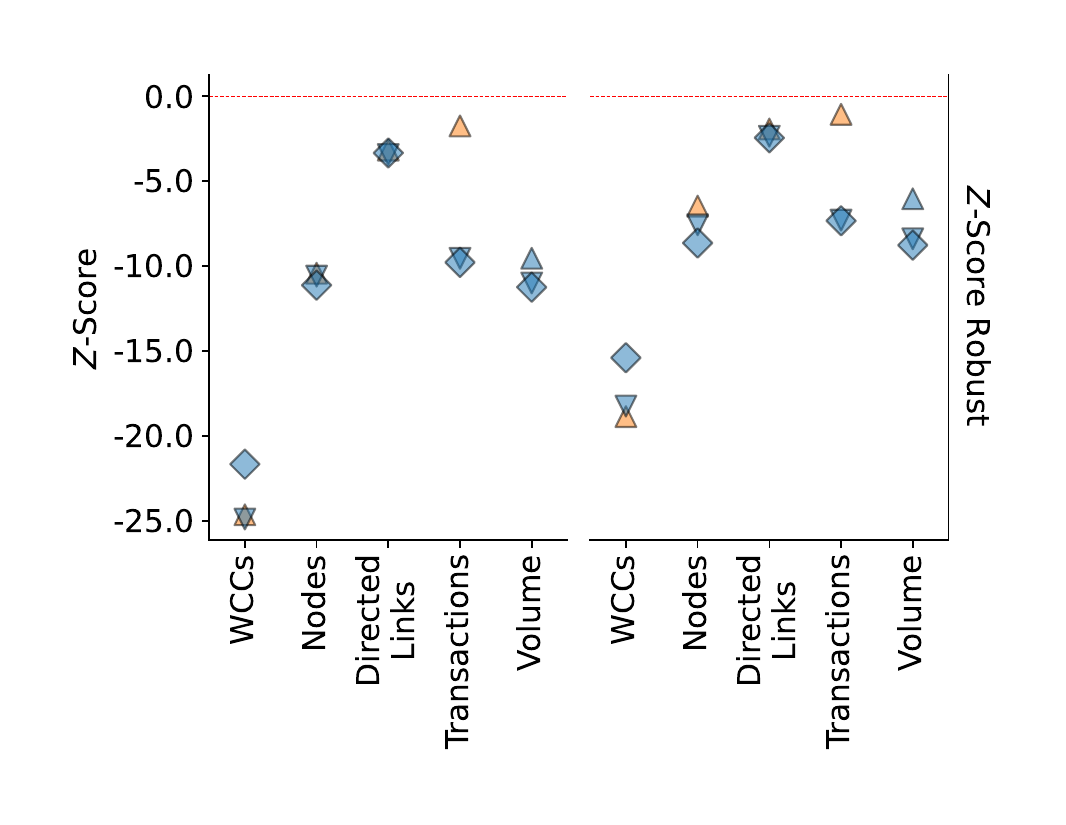}}
  \\
  \subfigure[Statistical significance for the \textbf{dagTin} category.]{\includegraphics[width=0.4\textwidth]{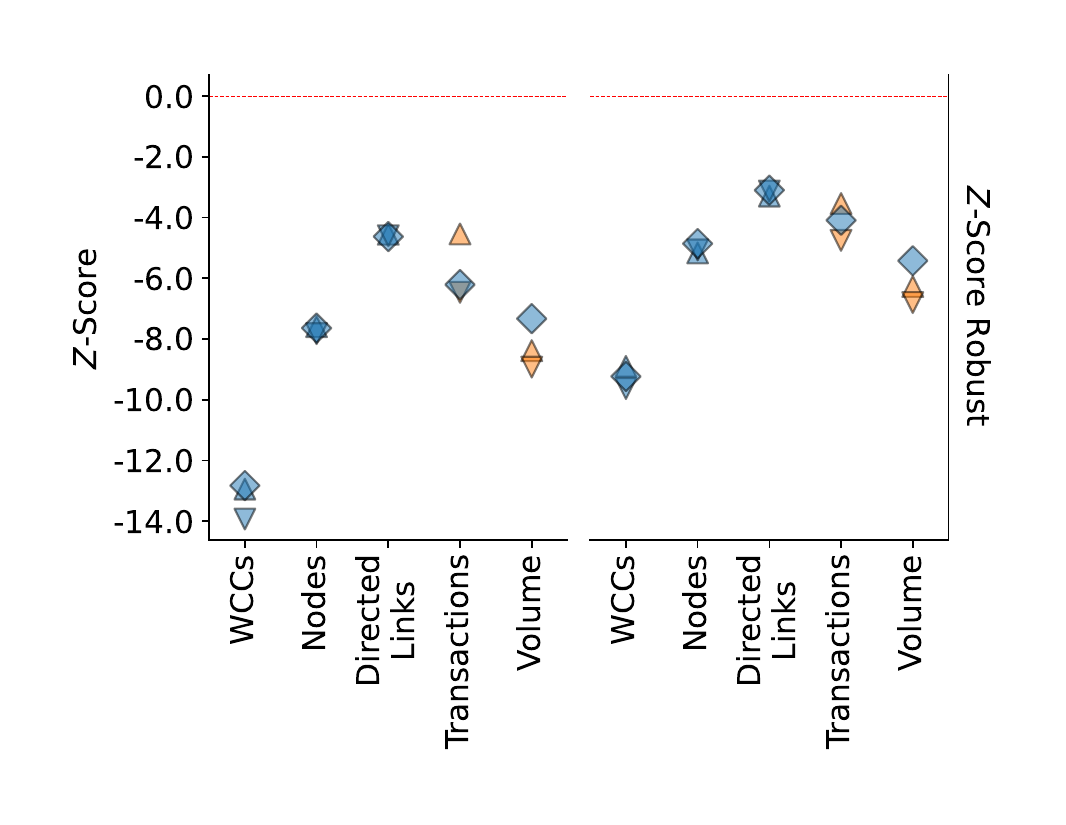}}
  \hspace{0.02\textwidth} 
  \subfigure[Statistical significance for the \textbf{dag0} category.]{\includegraphics[width=0.4\textwidth]{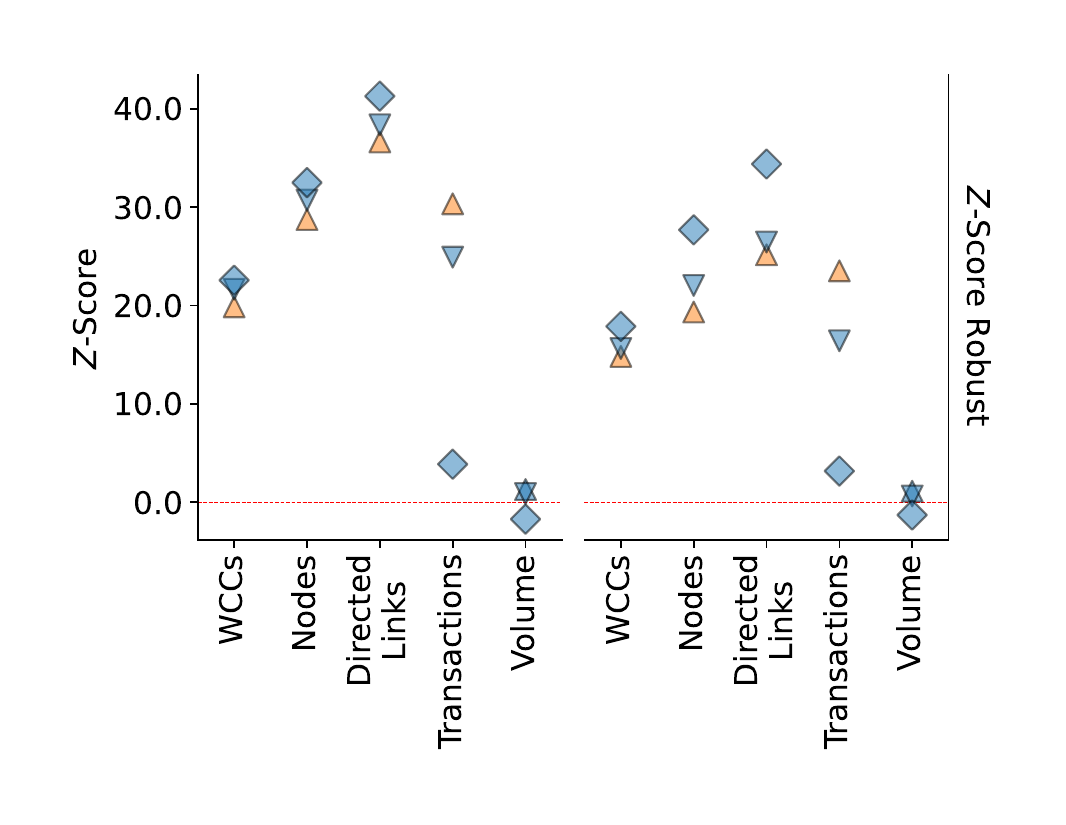}}
  \caption{Every feature has three markers, one per each null model used: "targets-swap" (downward triangle), "sources-swap", (upward triangle), and "both-swap" (rotated square). Both plots report the \textit{Z}-score and \textit{Z}-score robust (for further details see also Supplementary Material, Document A, Section "Significance"). The feature "WCCs" indicates the number of weakly connected components. The normality test is carried out by the Anderson-Darling test~\cite{stephens1979}: "yellow" markers represent values for which the p-value is less than 5\% (and therefore the normality hypothesis cannot be rejected).}
  \label{fig:topology_map_significance}
\end{figure}

The network features presented in Table \ref{tab:topological_groups_numb} are then compared to three types of null models (more details in Supplementary Material, Document A, Section"Significance"). In the first type of null models only the targets are swapped ("target-swap"), in the second only the sources ("source-swap"), and in the third either the source or target is swapped with a chance sampled from a uniform distribution ("both-swap"). When swapped, each directed link between two nodes carries with it all the transactions (and related data) between those two nodes. The comparison is carried out by using \textit{Z-}score and Robust \textit{Z-}score (for further details see Supplementary Material, Document A, Section "Significance"). The \textit{Z-}score represents the distance between the empirical value and the average value of the null models, but expressed in number of standard deviations. While, the Robust \textit{Z-}score represents the distance between the empirical value and the median of the null models, but expressed in number of inter-quartile ranges. Generally, the Robust \textit{Z-}score is considered when the normality assumption (necessary for the ordinary \textit{Z-}score) is rejected by the Anderson-Darling normality test \cite{stephens1979}. The results of this comparison with the null models are reported in Figure \ref{fig:topology_map_significance}. 

In this empirical network, there are 62 \textit{sccTmix} components. The largest strongly connected component falls in the category of \textit{sccTmix}. As explained before, a strongly connected component with suffix \textit{-mix} is both "receiving from" and "sending to" either a DAG and/or a single-node. In the null models, a large strongly connected component with mixed behaviour (\textit{sccTmix}) emerged, but it is also the unique strongly connected component. In other words, while in the empirical network there are 62 \textit{sccTmix} components, in null models there is only one which is also the largest. This result is not reported in Figure \ref{fig:topology_map_significance}(a) for obvious reasons: since the standard deviation and the inter-quartile ranges of the number of components (WCCs, weakly connected components) are equal to zero, both the \textit{Z-}scores and the Robust \textit{Z-}scores cannot be defined. 

Generally, the size of the largest strongly connected component in null models has the same size of the empirical one in terms of number of nodes. In fact, in Figure \ref{fig:topology_map_significance}(a) the \textit{Z-}score of the number of nodes is equal to zero. This means that the number of nodes in \textit{sccTmix} components of the empirical network is actually neither more nor less than expected in a random setting. Similarly, the number of directed links and transactions in \textit{sccTmix} components do not have a strong significant difference from the null models. In fact, the Robust \textit{Z-}score shows a value only between 2 to 6 times (of inter-quartile range) higher than the median value in null models. Nevertheless, the volume shows a much higher positive significance and the \textit{Z-}score is 14 to 25 times (i.e. standard deviations) higher than the average value of null models. To sum up, the presence of \textit{sccTmix} components in the empirical network is only significant (i.e. over-represented with respect to the null models) for its fragmentation into many different groups (WCCs) and for its volume.

The other types of strongly connected components (\textit{scc0}, \textit{sccTin}, \textit{sccTout}) are generally absent in the null models. This means that the presence of these types of strongly connected components cannot be explained only by randomness. The cyclic structures of those components may play a significant role in this economic network. The edges (\textit{edge\_scc2scc}) and nodes (\textit{bridge\_scc}) connecting different strongly connected components are also generally absent in the null models. This means that these particular topological components may reflect a significant functionality in this empirical network. As explained before, this could be related to the particular role that cycles play in real transaction networks for currency recirculation. 

In Figure \ref{fig:topology_map_significance}(e), the volume of \textit{dag0} components is not significant. The reason for this behaviour is partially explained in Section \ref{sec:results:dag_analysis}. In short, these components are mainly made by dyads and "collector" triads, where one user is collecting from other two. In fact, their composition could indicate trials among users not really interested in fully engaging with the rest of the network and therefore the exchanged volume is not more than random (more details in the Discussion Section \ref{sec:discussion}). Despite this, all the other features of \textit{dag0} components are positively significant. Since the normality assumption is rejected (except for "source-swap" models), the Robust \textit{Z-}score is generally more reliable for testing the statistical significance in this case. The number of components (WCCs) for \textit{dag0} is 15-20 times (i.e. inter-quartile ranges) higher in the empirical network than the median value of the null models. The number of nodes in \textit{dag0} is 20-25 times higher in the empirical network than the median value of the null models. The number of links is 25-30 standard deviations higher in the empirical network than the median value of the null models. However, the number of transactions has some variability which explains also the low significance of volume. To sum up, the number of \textit{dag0} components and their size (in terms of nodes and links) is more than expected from a random setting. This means that they may reflect a specific type of behaviour which characterises this network. However, the number of transactions and their volume is not generally more than expected from a random setting.

Finally, the presence of \textit{dagTin} (Figure \ref{fig:topology_map_significance}(d)), \textit{out-single-nodes} (Figure \ref{fig:topology_map_significance}(c)), and \textit{in-single-nodes} (Figure \ref{fig:topology_map_significance}(b)) components in the empirical network seems generally less than expected from the null models. In fact, the \textit{Z-}score has either a very high negative value or close to zero. A high negative value means that there is a statistically significant under-representation in the empirical network with respect to a random setting. 

For the \textit{in-single-node} category, the normality hypothesis cannot be rejected for "source-swap" and "target-swap" models, therefore the \textit{Z-}score is considered. The \textit{Z-}score of the number of components (WCCs) is about 15 times (i.e. standard deviations) less than expected on average in a random setting. The number of nodes is about 40 times less than expected, and the volume about 20 times less. The other features do no seem significant. 

In the \textit{out-single-node} category, only for the "source-swap" model the normality hypotesis cannot be rejected. For this reason, the Robust \textit{Z-}score is considered. The number of components (WCCs) is about 15-20 times (i.e. inter-quartile ranges) less than the median of the random setting. The number of nodes and the volume is 5-10 times less than the median value of the null models. Other network features do not show a strong significance.

For the \textit{dagTin} components, the normality hypothesis cannot be rejected for three features (i.e. WCCs, number of nodes, and directed links), so the Robust \textit{Z-}score is considered for those. The number of nodes are 10 times less than the median of a random setting, while nodes and directed links about 5 times. For the number of transactions and the exchanged volume, the normality hypothesis cannot be rejected for "source-swap" and "target-swap" models. The number of transactions is about 5 times less than median value, while the volume is about 9 times less. 

To sum up, the presence of \textit{dagTin}, \textit{out-single-nodes}, and \textit{in-single-nodes} in terms of number of components and nodes involved is surely less than expected in a random setting. This means that on a randomised network we can expect more of those components and in larger size. This result complements the previous finding on the presence of strongly connected components. In other words, a randomisation is expected to destroy cyclic structures and create more acyclic components and even single-nodes.

Concluding, the result suggests that the hypothesis on the particular role played by cyclic structures made in previous studies\cite{mattssonCirculation, iosifidis} could be confirmed by the statistically significant relative presence of strongly connected components and the significant relative absence of \textit{dagTin}, \textit{in-single-node} and \textit{out-single-node}. In Section \ref{sec:discussion}, the presence of \textit{in-single-nodes} could be hypothetically associated to the creation of "fake" accounts. Surprisingly, the large positive \textit{Z-}scores (or over-representation) for \textit{dag0} may indicate another type of behaviour, where small groups of people are either trying out the system or collecting Sarafu from each other without further engagement with the economic network (see Section \ref{sec:discussion} for more details). In the next two sections \ref{sec:results:one_tx_users} and \ref{sec:results:dag_analysis}, we will try to explain how the relative significant presence of \textit{dag0} and the relative significant absence of \textit{dagTin}, \textit{out-single-node}, and \textit{in-single-nodes} can be indeed associated to specific behavioural dynamics. For instance, these components are strongly associated to one-time users (see Section \ref{sec:results:one_tx_users}). Furthermore, \textit{dag0} and \textit{dagTin} are related to the presence of dyads and "collector" triads (i.e. one user collecting from other two)(see Section \ref{sec:results:dag_analysis}). These findings will be also compared to a recent qualitative study\cite{gaming_sarafu2024} and discussed in Section \ref{sec:discussion}.

\subsection{One-Time Users}  \label{sec:results:one_tx_users}

Many users made only one transaction during the observation period. Our findings suggest that this phenomenon is closely related to the topological categories presented in the previous section~\ref{sec:results:top_cat}. In Table \ref{tab:one_tx_users_top}, such \textit{One-Time} users are grouped by topological categories. Since they were involved in one operation in only one direction (receiving or sending), these users can belong only either to acyclic components (DAGs) or single-nodes. In particular, 47\% of \textit{in-single-nodes}, 44.6\% of \textit{dagTin} users, 78.5\% of \textit{dag0} users, and 61.9\% of \textit{out-single-nodes} took part only in one operation, either as receivers or as senders (see Table~\ref{tab:topological_groups_numb} for comparison). 

One of the criticisms that Sarafu got in this period was its use of rewards for bringing/recommending a new user\cite{gaming_sarafu2024} (more details in Section~\ref{sec:discussion}). An "old" user could recommend a "new" user's phone number to get a bonus after its registration. If the "old" and the "new" users agreed upon it, the "old" user could also forward to its account the initial disbursement that the "new" user received from Grassroots Economics. In fact, no previous agreement between the "old" and the "new" users was necessary, as far as the "old" user had access to that phone number the operation would have been successful. In simple words, this means that some existing users could get the reward and the initial disbursements by recommending and then registering "fake" accounts using phone numbers of other people, consensually or not. Analysing the transaction network, this operation would be recorded as an operation involving one-time users sending Sarafu only once to transfer this initial disbursement (or part of it) to at least one other user (i.e. the "old" user). In fact, these one-time users which sent Sarafu and then stopped are 22.7\% of the total number of registered users, but their transactions correspond only to 1.2\% of the total volume. It is not possible to accurately estimate the size of this phenomenon, but at least it is possible to define broadly its boundaries. 

The presence of one-time users seems to strengthen the suspicion of use of multiple "fake" accounts. Nonetheless, the main topological groups corresponding to this type of activity would be identified mainly as \textit{in-single-nodes} and \textit{dagTin} (moderate green and soft blue nodes, in Figure \ref{fig:topology_subgraph}). In both these categories, we can observe one node in a \textit{sccTmix} component which is receiving from one or more nodes outside of that component. The node sending can either be isolated (\textit{in-single-node}) or connected with other nodes \textit{dagTin}. For instance, the DAG could include one node collecting from other nodes, or a chain of nodes transferring to each other. One-time users falling into the category of \textit{in-single-nodes} exchanged an amount of volume equals to 1\% of the total. While one-time users falling into the category of \textit{dagTin} exchanged an amount of volume equals to 0.1\% of the total. Therefore, even though the phenomenon can be observed in the data, its size may be still negligible. 

Another important result to remark is that the majority of \textit{dag0} users (78.5\%) sent or received only one transaction. Moreover, there are 289 transactions which were made by a one-time sender towards a one-time receiver, these transactions are happening within \textit{dag0} components and are equal to 31.9\% of the total number of \textit{dag0} transactions. In the previous section, it was observed that the presence of \textit{dag0} components is more than random. This reinforces the presumption that this category may indicate mostly isolated pairs or group of users simply trying out or testing the system without really engaging with the rest of the network. A behaviour which we could expect happening very frequently in a novel complementary digital currency like this one.

Concluding, in the previous section we observed that the presence of \textit{dagTin}, \textit{in-single-nodes}, and \textit{out-single-nodes} is generally less than random. In this section, we observed that about half of the users falling into these categories used the system only one time. In particular, users in \textit{dagTin} and \textit{in-single-nodes} could be related to "fake" accounts used to collect currency. Nonetheless, this phenomenon described also in a recent qualitative study\cite{gaming_sarafu2024} have a negligible size (see Section \ref{sec:discussion} for more details). Furthermore, in the previous section, it was observed that the presence of \textit{dag0} components is positively significant, but the majority of their users used the system only one time. This reinforces the suspicion that users in \textit{dag0} components were only trying out the system. In the next section, we explore the three-nodes motifs occurring in the acyclic components (DAGs) with a particular focus on \textit{dag0} and \textit{dagTin} components.

\begin{table}[!ht]
\centering
\begin{tabularx}{\textwidth}{|p{0.15\linewidth}|X|X|X|X|}
\hline
\textbf{Topological Category} & \textbf{Users with One Outgoing Transaction} & \textbf{Users with One Incoming Transaction} & \textbf{Outgoing Volume} & \textbf{Incoming Volume} \\ \hline
in-single-node  & 7,011 & - & 1,886,274.21  & - \\ \hline
dagTin       & 1,109 & 70 & 241,881.0  & 5,542.0 \\ \hline
dag0         & 619 & 327 & 53,439.0   & 42,820.0 \\ \hline
out-single-node  & - & 746 & - & 39,789.03    \\ \hline
dagTmix       & 132 & 10  & 35,161.0 & 486.0  \\ \hline
dagTout       & 113 & 26 & 14,601.0  & 3,647.0   \\ \hline 
total         & 8,984 & 1,179 & 2,231,356.21 & 92,284.03 \\ \hline
\end{tabularx}
\caption{Users who took part only in one operation (One-Time Users), either as a receiver or a sender only.}
\label{tab:one_tx_users_top}
\end{table}

\subsection{Triads}  \label{sec:results:dag_analysis}

In this section, a triadic census analysis of the acyclic components (DAGs) is provided. It is possible to observe that 90.6\% of DAGs have a size between 2 and 5 (see Table S2 in Supplementary Material, Document A, Section "Topology"). Excluding the DAGs made by one-time users, the rest of them counted 1,571 nodes with 1,115 directed links, 3,345 transactions and a total volume of 510,945.13 Sarafu (0.2\% of the total volume). In acyclic components, the majority of transactions among users who used the system more than once belong to the \textit{dagTin} category (see Figure \ref{fig:scc_dag_size}(a)). 

Since the majority of DAGs are dyads and triads, in this section a triadic census analysis is implemented by using one of the most common existing nomenclatures~\cite{triadic_census}. However, only 4 of those 16 types of triads are found to have a statistically significant presence in this network: 012, 021C, 021U, and 021D. The dyadic triad 012 is a simple dyad (from A to B). The triad 021C is a “brokerage” interaction (from A to C through B - where B is the "broker"). The triad 021U represents one central user collecting the currency of the other two (from A to B, from C to B - where B is the "collector"). The triad 021D represents one central user sending to two other users (from B to A, from B to C - where B is the "distributor").

\begin{figure}[!ht]
  \centering
  \subfigure[Statistical significance of triad 021U for \textbf{dag0} and \textbf{dagTin}.]{\includegraphics[width=0.4\textwidth]{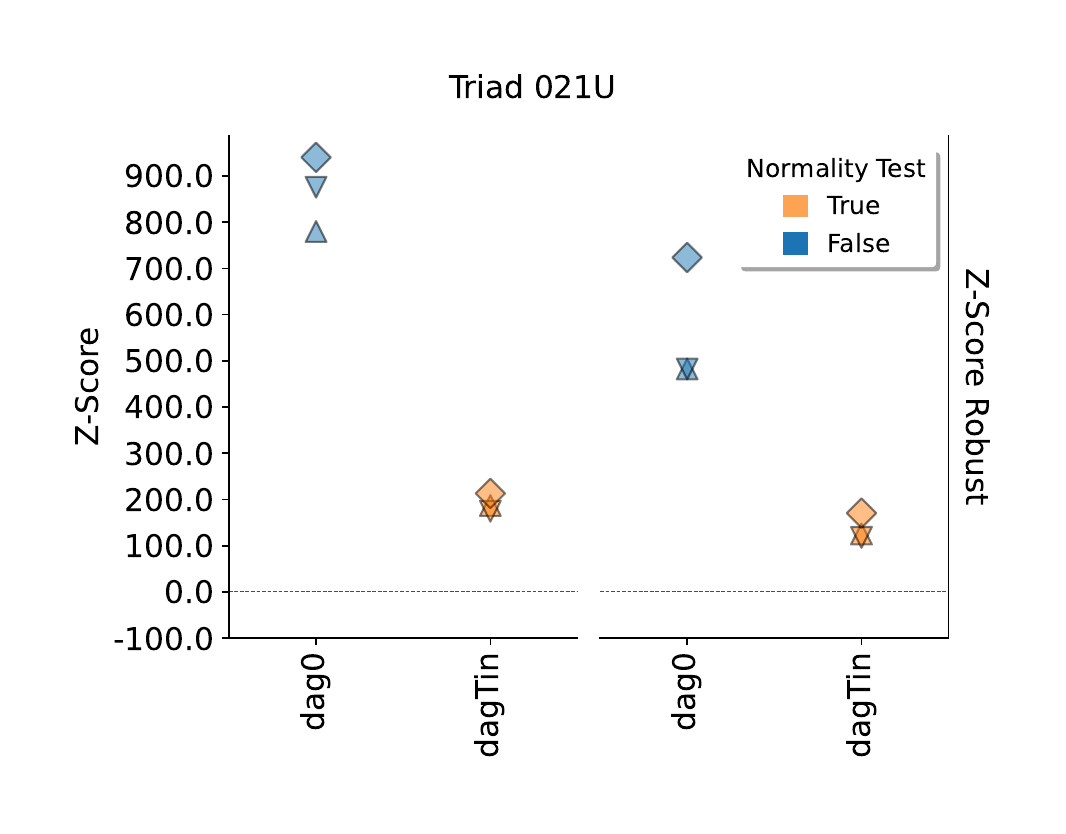}}
    \hspace{0.02\textwidth}
  \subfigure[Statistical significance  of triad 021U for \textbf{dagTmix} and \textbf{dagTout}.]{\includegraphics[width=0.4\textwidth]{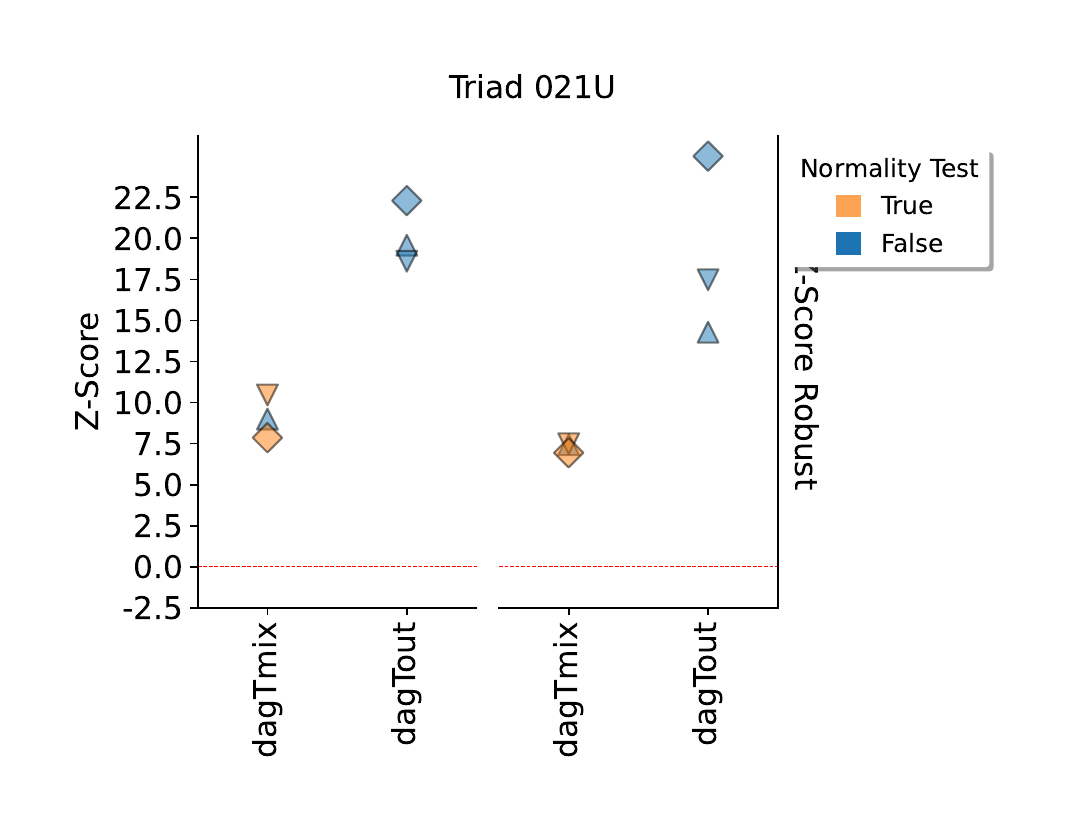}}
  \\
  \subfigure[Statistical significance of triads 021C, 012, 021D for \textbf{dag0}.]
  {\includegraphics[width=0.4\textwidth]{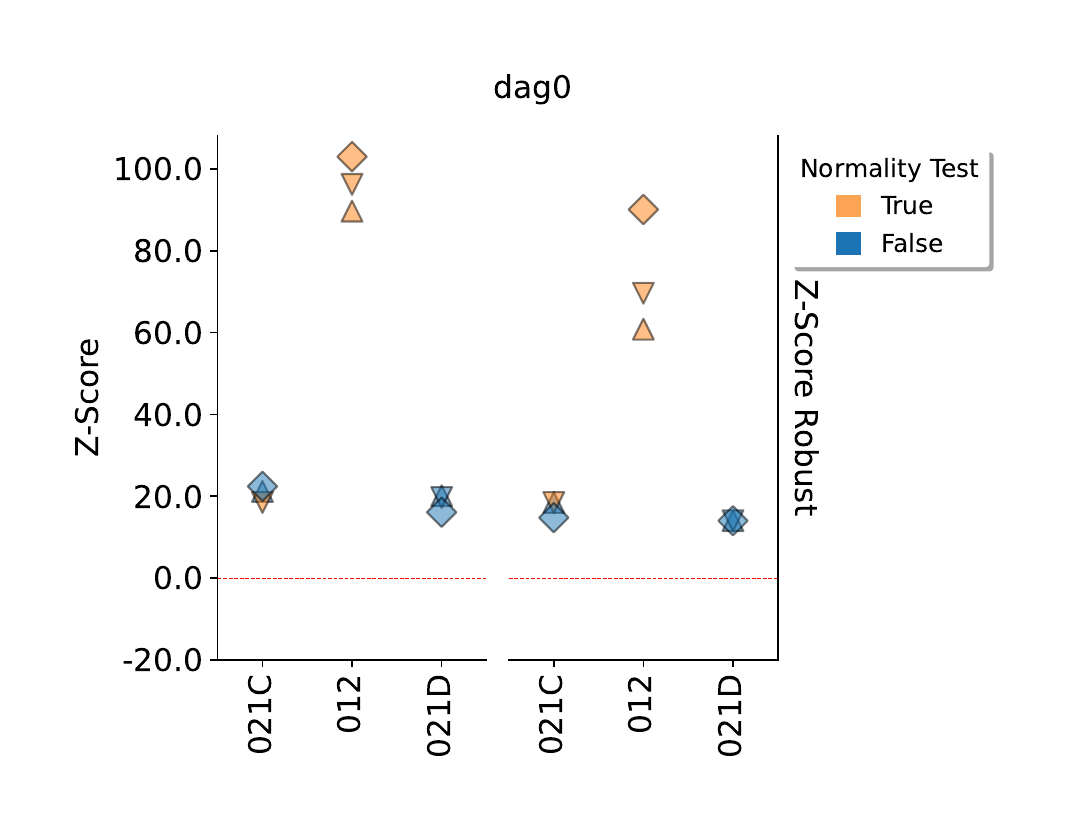}}
  \caption{On the top, the statistical significance of triad 021U for each type of DAG. The triad 021U represents one central user collecting the currency of other two. On the bottom, the statistical significance for triads 021C, 012, and 021D for "dag0" components. The normality test is carried out by the Anderson-Darling test ~\cite{stephens1979}: "yellow" dots represent values for which the p-value is less than 5\% and therefore the normality hypothesis cannot be rejected. The three markers correspond to one of the of null models used for this test: "targets-swap" (downward triangle), "sources-swap", (upward triangle), and "both-swap" (rotated square) - for further details see also Supplementary Material, Document A, Section "Significance".}
  \label{fig:combined_triads_dag}
\end{figure}

The "collector" triad 021U is the most significant type of triad across all the DAGs (see Figure~\ref{fig:combined_triads_dag}). For \textit{dag0}, the Robust \textit{Z-}score is up to 700 time higher than the median value in null models (in terms of inter-quartile ranges). For \textit{dagTin}, the \textit{Z-}score is up to 200 time higher than the average value in null models (in terms of standard deviations). From the previous sections, we know that the majority of \textit{dag0} users and half of \textit{dagTin} used the system only one-time. This is coherent with the presence of "collector" users reported here: an active user was probably collecting from other "fake" accounts which made only one operation. While \textit{dagTin} components have at least one user sending to a strongly connected component, \textit{dag0} components are isolated. This means that the "collector" user in a \textit{dag0} component was simply only hoarding the currency.

Nonetheless, it is worth mentioning the statistical significance of triads 021C, 012, and 021D in \textit{dag0} components. In fact, all these triads seems to be also particularly prominent in \textit{dag0} components, especially isolated dyads 012. Since the majority of \textit{dag0} are one-time users, it is fair to assume that these are groups of people simply trying out the system. Even if statistically significant the size of these phenomena is not so relevant to compromise the whole system. In fact, the volume of \textit{dag0} is only the 0.002\% of the total volume of Sarafu (see Table ~\ref{tab:topological_groups_numb}). Finally, the other triads in \textit{dagTin}, \textit{dagTout}, and \textit{dagTmix} have a very low or even negative \textit{Z-}scores. This means that there is no particular triad formation playing a meaningful role in the flow of Sarafu in those topological categories, except for 021U (the "collector" triad).

Concluding, the presence of "collectors" is highly significant in acyclic components and are identifiable by the triad 021U. As expected, most of the "collectors" lay in \textit{dagTin} and \textit{dag0} components: 44.6\% of \textit{dagTin} accounts and 78.5\% of \textit{dag0} accounts were used only once. This evidence increases the suspicion that \textit{dagTin} identify users collecting Sarafu from "fake" accounts in at least half of the cases, while \textit{dag0} identify users simply trying out the system in the majority of the cases. In the next section, the dynamics of circulation are explored. The intent is to observe differences in circulation between cyclic (SCCs) and acyclic components (DAGs) in line with the previous findings.

\subsection{Recirculation}  \label{sec:results:recirculation}

We describe the circulation by using a temporal metric, the \textit{time of recirculation}. A recirculation operation is the time difference between the first of all the incoming transactions and the last of all the outgoing transactions, before the next incoming transaction arrives. In Figure \ref{fig:recirculation_illustration}, the concept of recirculation operation is explained. All the transactions happening between time \textit{t} and time \textit{t+3} are included in the recirculation operation. As soon as an outgoing transaction is followed by another incoming transaction (at time \textit{t+4}), the recirculation operation gets closed. This implies that the same node can have many recirculation operations. In other words, for each set of incoming transactions followed by a set of outgoing transactions, a recirculation operation is recorded for that particular user.

In total, there were 123,741 recirculation operations. The total amount of transactions included in these operations was of 328,191 transactions, which are equal to 91.1\% of the total number of transactions and 94\% of the total volume. As described before, one recirculation operation can be made of several outgoing and incoming transactions. Almost half (49.4\%) of those transactions took place among recirculating users. In other words, the outgoing transaction of one recirculating user became the incoming transaction of another user who recirculated afterwards in almost half of the cases. There were 9,984 recirculating users (25.31\% of the total), but they involved up to 23,488 other non-recirculating users, either as only-senders or only-receivers of these recirculating operations. In short, it is remarkable to observe that 25.31\% of users were responsible for 91.1\% of transactions (94\% of the total volume) and recirculating almost half of the transactions among them (49.4\%). Additional information on the statistical significance of recirculation is provided in Supplementary Material (Document A, Section "Recirculation").

In terms of recirculation operations, it is remarkable the role played by \textit{in-single-nodes} (Table \ref{tab:recirculation_tx}). The \textit{in-single-nodes} are the second biggest category of users who initiated recirculation operations. This means that 5.8\% of the total transactions involved in recirculation operations was made by single-nodes sending to SCCs. Note also that this is true across all temporal categories, but decreasing from the slowest (LFQ3) to the fastest (HFQ1). This means that the role of users in \textit{in-single-nodes} was that one of providing liquidity to other users in strongly connected components (\textit{sccTmix} and \textit{sccTin})) which eventually recirculated it, after a few days (the majority) or a few seconds. This may confirm the suspicion that \textit{in-single-nodes} were mostly made of "fake" accounts created with the purpose of collecting Sarafu to re-use it afterwards. Especially, considering that half of them made only one operation, as explained in Section \ref{sec:results:one_tx_users}).

In Table \ref{tab:freq_topology}, the distribution of users for the main topological groups across the largest temporal categories is reported. The temporal categories are defined in Section~\ref{sec:methods}. Most of recirculation operations are concentrated in three main temporal categories: users with only recirculation activity longer than a day (LFQ3), users with recirculation activity less than 20 minutes (HFQ1), and users with recirculation activity across all the temporal categories (HFQ1, HFQ2, HFQ3, LFQ3). Users belonging to strongly connected components have the most diverse temporal behaviour, and most of recirculation is happening within SCCs: 91.99\% of recirculating users belong to \textit{sccTmix} components, 2.28\% to \textit{sccTin}, 1.49\% to \textit{scc0}, and 0.5\% to \textit{sccTout} components. This also confirms the role played by cyclic structures in the recirculation of a currency system. On the other hand, most of users in DAGs do not report such temporal diversity and they report mostly either very high frequency of recirculation (HFQ1) or very low (LFQ3).

\begin{table}[!ht]
    \centering
    \begin{tabular}{|l|l|l|l|l|l|}
    \hline
        \textbf{} & \textbf{HFQ1} & \textbf{HFQ2} & \textbf{HFQ3} & \textbf{LFQ3} & \textbf{Tot.Top.Cat.} \\ \hline
        \textbf{sccTmix} & 64,096 & 100,646 & 109,976 & 128,947 & 403,665 \\ \hline
        \textbf{in\_single\_node} & 439 & 2,548 & 3,247 & 14,746 & 20,980 \\ \hline
        \textbf{sccTin} & 933 & 1,172 & 954 & 1,043 & 4,102 \\ \hline
        \textbf{edge\_dag2scc} & 79 & 333 & 461 & 2,143 & 3,016 \\ \hline
        \textbf{edge\_scc2scc} & 48 & 129 & 187 & 938 & 1,302 \\ \hline
        \textbf{scc0} & 323 & 174 & 250 & 514 & 1,261 \\ \hline
        \textbf{sccTout} & 181 & 139 & 182 & 490 & 992 \\ \hline
        \textbf{Tot.Freq.Cat.} & 66,283 & 105,474 & 115,462 & 1,495,33 & 436,752 \\ \hline
    \end{tabular}
     \caption{Number of transactions involved in recirculation operations. Note that the total is greater than 328,191. This is due to the fact that 108,561 were transactions from and to recirculating users. For this reason, they are counted twice (i.e. first, as "incoming", and then, as "outgoing"). The table only reports the results for the first 7 topological groups. A complete version of the table can be found in Supplementary Material B.}
    \label{tab:recirculation_tx}
\end{table}

\begin{table}[!ht]
    \centering
    \begin{tabular}{|p{0.06\textwidth}|c|c|c|c|c|p{0.07\textwidth}|}
    \hline
        \textbf{} & \textbf{scc0} & \textbf{sccTin} & \textbf{sccTout} & \textbf{sccTmix} & \textbf{dagTin} & \textbf{Tot.Freq. Cat.} \\ \hline
        \textbf{HFQ1} & 70 & 46 & 12 & 967 & 13 & 1113 \\ \hline
        \textbf{LFQ3} & 47 & 108 & 20 & 1682 & 46 & 1926 \\ \hline
        \textbf{*All} & 6 & 22 & 8 & 2092 & 0 & 2128 \\ \hline
        \textbf{Tot.Top. Cat.} & 216 & 334 & 82 & 9185 & 120 & 9984 \\ \hline
    \end{tabular}
    \caption{Number of users involed in recirculation operations. The users are counted only for the main topological group and the three largest temporal categories (HFQ1, LFQ3, and All). A complete version of the table can be found in Supplementary Material B. The percentages per each topological group represent the share of users for each temporal category. The category "All" reports users who were active in all of the considered temporal categories (HFQ1, HFQ2, HFQ3, LFQ3).}
    \label{tab:freq_topology}
\end{table}

Concluding, recirculation took place among 25.31\% of users who were responsible for 91.1\% of transactions (94\% of the total volume). Almost half of those transactions was recirculated only among recirculating users (49.4\%). Moreover, 75\% of those transactions was recirculated within less than two days. The majority of recirculating users belong to strongly connected components, 91.99\% only to \textit{sccTmix} components. Furthermore, \textit{in-single-nodes} played an important role in initiating about 5.8\% of recirculating operations. Since almost half of those accounts was used only once, this means that \textit{in-single-nodes} may indeed identify accounts which were used only to collect Sarafu. This liquidity was eventually injected and recirculated in the rest of the economy through strongly connected components. In the next Section \ref{sec:discussion}, the results are summarised and analysed to appreciate the complete picture of these findings.

\section{Discussion} \label{sec:discussion}

A recent study on Sarafu network argued that the adoption of community currency system is more efficient and effective than the usual cash transfer program using fiat currency~\cite{ussher_complementary_2021}. In that paper, it is argued that the use of local currency stimulates the local economy, while providing humanitarian aid. However, the use of "Sarafu token", as cash transfer program which was in place between 2020 and 2021, recently also received some criticisms~\cite{gaming_sarafu2024, barinaga_2020}. Some of those criticisms are discussed in this section. In the most recent qualitative study~\cite{gaming_sarafu2024}, the findings from 31 interviews and 6 focus groups (of 8-12 participants) suggested that some users may have tried to "game" the system by taking advantage of its rewarding and cashing-out programs. In that work, the size of the phenomenon is not further investigated, but four different \textit{gaming} strategies are described and reported here below.

\begin{itemize}

\item \textbf{Strategy N.1}. An existing account could register more than one phone number. Every new phone number is associated to a new account which will automatically get an initial disbursement. In principle, by having direct physical access to this new phone number, a first user could send to itself this initial disbursement to accumulate more Sarafu. Moreover, from this new phone number it would be possible to communicate the phone number of someone else from whom it was invited to join the system (i.e. recommendation system). In this way, the first user could get also a reward (in Sarafu) for having brought this new member in the network. This recommendation system is also described in a data descriptor paper~\cite{mattssonSarafu} .

\item \textbf{Strategy N.2}. Similarly to \textit{Strategy N.1}, an existing user could register the phone number of other people in a non-consensual way. The user, who had access to someone else phone for enough time, could grant itself all the benefits described in \textit{Strategy N.1}.

\item \textbf{Strategy N.3}. Users could fake transactions to get some promotional bonus and avoid the (weekly, and then, monthly) holding fee. The holding fee (or \textit{demurrage} charge) is also described in the data descriptor paper~\cite{mattssonSarafu}. On the other hand, the functioning of the promotional bonus is not officially described. The disbursement of the promotional bonus took place four times in the analysed period~\cite{mattssonSarafu}. The qualitative study~\cite{gaming_sarafu2024} reports a specific promotional strategy which seems to be connected to a reward for triads formation: \textit{"A buys from B, C and D, who in turn trade with E, F and G"}(p.44 ~\cite{gaming_sarafu2024}), and \textit{"I send him, he sends to her and she sends to me"}(p.46 ~\cite{gaming_sarafu2024}). The first triad formation strategy seems to describe a "brokerage" triad (021C). The second triad formation strategy seems to describe a cycle of length 3 (030C).

\item \textbf{Strategy N.4}. Every saving group (or \textit{Chama}) registered in the Sarafu system was entitled to a partial/limited cash-out (i.e. exchange Sarafu for Kenyan Shillings). A user who wanted to exchange back its own Sarafu to Kenyan Shilling could potentially register to many different savings groups to escape the cash-out limitations.

\end{itemize}

In this section, the first three \textit{gaming} strategies are analysed by considering the transaction data for the same period. In fact, the integration of qualitative and quantitative methods could help reconstructing a complete picture of this phenomenon. In the next paragraphs, we will try to match the quantitative findings presented in this paper with the description of these "gaming" strategies just mentioned.

The strategies \textbf{N.1} and \textbf{N.2} are easy to detect by looking at the topology of the transaction network. The above-mentioned behaviours imply the presence of one existing account connected with other accounts which only send Sarafu to it and then stay mostly inactive. As observed in Section \ref{sec:results:one_tx_users}, about half of the users falling in the categories of \textit{in-single-node} and \textit{dagTin} used the system only once, while the majority of users in \textit{dag0} used the system only one time. Furthermore, in \textit{dagTin} there is a significant presence of "collectors" (Section \ref{sec:results:dag_analysis}). Strategies \textbf{N.1} and \textbf{N.2} can be therefore confirmed by the data. However, one-time users in \textit{in-single-nodes}, \textit{dag0}, and \textit{dagTin} moved overall only 1.19\% of the total volume (see Table \ref{tab:one_tx_users_top}). To sum up, the quantitative analysis confirms the strategies \textbf{N.1} and \textbf{N.2} described above, but the size of the phenomenon does not seem big enough to compromise the rest of the system. 

The strategy \textbf{N.3} describes triad formation through "fake" transactions with the only intent to accumulate Sarafu and eventually cashing them out (strategy \textbf{N.4}). According to the same qualitative study (pp. 46-47~\cite{gaming_sarafu2024}), some users were meeting regularly with the intention of "faking" transactions to escape the holding fee and trying unlocking some promotional bonus. If the intent was getting Sarafu and cashing it out for Kenyan Shillings (Strategy \textbf{N.4}), then we would observe this behaviour in DAGs. On the other hand, if the intent was getting Sarafu to eventually participate in the economic network, we would observe this behaviour in SCCs. Nonetheless, while "faking" for cashing-out was an unintended consequence of the currency design, the second type of behaviour was somehow intended in the currency design and indistinguishable from ordinary transactions happening in the Sarafu economic network (or SCCs). 

The exploration of strategy \textbf{N.3} in DAGs leads us to the results presented in Section \ref{sec:results:dag_analysis}, where a triadic census analysis of DAGs was carried out. The results suggest a significant presence of dyads (012) and many other different types of triads in \textit{dag0} components. However, since 78.5\% of users in those components made only one transaction, it is difficult to relate \textit{dag0} with this type of strategy. On the other hand, 55.4\% of users in \textit{dagTin} made more than one transaction and the "collector" triad (021U) is the most significant one in there. This means that \textit{dagTin} components are a good candidate for the description of strategy \textbf{N.3}, if the triad formation would not imply a cycle. Excluding the volume of one-time users operations, the resulting amount of currency exchanged in \textit{dagTin} components is equal to about 0.01\% of the total volume. 

The exploration of strategy \textbf{N.3} in SCCs can be also addressed by the results in \ref{sec:results:recirculation}. In that section, it is reported that 91.1\% of recirculating operations was happening within \textit{sccTmix} components. However, 15.8\% of the transactions in \textit{sccTmix} components composing those recirculating operations were happening within about 20 minutes (HFQ1)(in Table \ref{tab:recirculation_tx}). Moreover, 11.15\% of users engaged only in high-frequency operations (HFQ1) and the majority of them (86.9\%) laying in \textit{sccTmix} components (in Table \ref{tab:freq_topology}). In other words, there is indeed the suspicion that high frequency recirculating operations (HFQ1) happened with the purpose of unlocking some reward or escaping the holding fee, as described by strategy \textbf{N.3}. However, these rewards were also spent afterwards in Sarafu economic network (or SCCs), as initially planned by the currency designers.

Concluding, the strategies \textbf{N.1}, \textbf{N.2} and \textbf{N.3} can be indeed identified by analysing the behaviour of users in \textit{dag0}, \textit{dagTin}, \textit{in-single-node}, and \textit{sccTmix} components. Strategies \textbf{N.1} and \textbf{N.2} can be related to one-time users in \textit{in-single-nodes}, \textit{dag0}, and \textit{dagTin} who moved overall 1.19\% of the total volume. Strategy \textbf{N.3} can be related to triad formation. If the triad formation was happening with the intent of unlocking rewards to cash-out later, then we would observe a significant presence of triads in DAGs. Otherwise, if the triad formation was happening with the intent of unlocking rewards to engage with the rest of the economy, then we would observe a significant presence of triads in strongly connected components. In the first case, the presence of "collector" triads \textit{dagTin} components could indeed confirm this behaviour (0.01\% of the total volume). In the second case, we observed that 11.15\% of users were engaging only in high frequency operations (\textit{HFQ1}, less than 20 minutes) and the majority (86.9\%) of them are part of \textit{sccTmix} components. However, a triadic census on strongly connected components cannot reveal further details, since those users were embedded with the rest of the economy.

These findings answer also the research questions presented at the beginning of this paper. The identified topological categories are shown to be relevant for the study of a payment system (\textbf{RQ1}). Since they relate with the circulation of currency, these topological categories succeeded in identifying different temporal behaviours within the Sarafu economic network (\textbf{RQ2}). Moreover, these techniques were also used to describe different levels of engagement in a transaction network (\textbf{RQ3}). This shows the importance of studying the structure and the dynamics of an economic network to minimise the risk of unexpected outcomes in monetary interventions. This work also added further nuances on the study of circulation in currency networks. In particular, the role of cyclic structures presented in previous works \cite{iosifidis, mattssonCirculation} is confirmed by the meaningful differentiation between cyclic components (SCCs) and acyclic components (DAGs and single-nodes). Finally, as observed by a previous quantitative analysis of this data~\cite{cooperative_sarafu_2023, temporal_sarafu_2022}, the Sarafu token network generally succeeded in stimulating the local economy during the COVID-19 crisis by engaging the majority of its users within the local economic network.


\section*{Acknowledgements}

The Author thanks his supervisor Prof. J\'anos Kert\'esz for his advise. Thanks are due to the financial support by Freiburg Institute For Basic Income Studies.

\section*{Additional information}

The author declares no competing interests.

\section*{Data availability} 

The Sarafu data 2020-2021~\cite{ruddick_sarafu_2021} is available for download at UK Data Service (UKDS) under End User License (\url{https://reshare.ukdataservice.ac.uk/855142/}) after registration. A data description paper is also available for download~\cite{mattssonSarafu}.

\section*{Software availability} All software used in this study are available under an open-source licence: 
\begin{itemize}
  \setlength\itemsep{0em}

    \item \texttt{networkx v.3.1.}~\cite{networkx_2008}
    \item \texttt{scipy v.1.9.1.}~\cite{SciPy-NMeth}
    \item \texttt{numpy v.1.23.0}~\cite{harris2020array}
    \item \texttt{powerlaw v.1.5}~\cite{powerlaw2014}
    \item \texttt{seaborn v.0.11.2}~\cite{waskom_seaborn_2021}
    \item \texttt{matplotlib v.3.5.2}~\cite{hunter_matplotlib_2007}
    \item \texttt{pandas v.1.4.4.}~\cite{reback_pandas-devpandas_2022}
    \item \texttt{pycirclize v.1.4.0}~\cite{pyCirclize}
    \item \texttt{gephi v.0.10}~\cite{gephi}

\end{itemize}

\section*{Code availability}
The code required to construct, randomise, and analyse the network is included the \textbf{Supplementary Material Document B}. The Supplementary Material Document B includes the following files:
\begin{itemize}
    \setlength\itemsep{0em}
    \item \textbf{File 1} Notebook used for cleaning the data and graph randomisation
    \item \textbf{File 2} Notebook used for analysis
    \item \textbf{File 3} Notebook used for the creation of Figures and Tables
    \item \textbf{File 4} Python file with functions used for the randomisation 
    \item \textbf{File 5} Python file with functions used for the analysis and visualisation 
\end{itemize}
In the \textbf{Supplementary Material Document A}, there is additional information which were also referenced in this text. The Supplementary Material Document A and Document B are available for download.

\bibliography{sample}

\end{document}